\documentclass[12pt]{article}
\pdfoutput=1
\usepackage{amsmath,amssymb,amsthm,amsxtra,overpic,bbm,bm,epsfig,subfigure}
\usepackage{hyperref}
\usepackage{graphicx}
\usepackage{color}
\usepackage{comment}
\usepackage{epstopdf}
\usepackage{float}
\numberwithin{equation}{section}
\usepackage{cite}
\usepackage{multirow}
\usepackage{hyperref}
\usepackage{slashed,stmaryrd}

\usepackage{lscape}%
\usepackage{array}
\usepackage{booktabs}%

\def\thefootnote{\fnsymbol{footnote}}

\begin{document}

\vspace{0.2cm}

\begin{center}
{\Large\bf Lepton Flavor Mixing and CP Violation in the Minimal Type-(I+II) Seesaw Model with a Modular $A^{}_4$ Symmetry}
\end{center}

\vspace{0.2cm}

\begin{center}
{\bf Xin Wang}~$^{a,~b}$~\footnote{E-mail: wangx@ihep.ac.cn},
\\
\vspace{0.2cm}
{\small
$^a$Institute of High Energy Physics, Chinese Academy of Sciences, Beijing 100049, China\\
$^b$School of Physical Sciences, University of Chinese Academy of Sciences, Beijing 100049, China}
\end{center}

\vspace{1.5cm}

\begin{abstract}
In this paper, we study the implications of the modular $A^{}_4$ flavor symmetry in constructing a supersymmetric minimal type-(I+II) seesaw model, in which only one right-handed neutrino and two Higgs triplets are introduced to account for the tiny neutrino masses, flavor mixing and CP violation. We consider the most economical case where the right-handed neutrino and the Higgs triplets in this model are assigned into the trivial one-dimensional irreducible representation of the modular group $A^{}_4$, and all the modular forms are with the lowest weights they can take. We find that the octant of the mixing angle $\theta^{}_{23}$ strongly depends on the hierarchy of free model parameters in the charged-lepton sector. We also show that the neutrino mass matrix can possess an approximate $\mu-\tau$ reflection symmetry for some specific values of free parameters. Moreover, our model predicts relatively large masses of three light neutrinos, thus can be easily tested in future neutrino experiments.

\end{abstract}


\def\thefootnote{\arabic{footnote}}
\setcounter{footnote}{0}

\newpage

\section{Introduction}\label{sec:intro}

Neutrino oscillation experiments in the past two decades have provided us with the very solid evidence that neutrinos are massive and lepton flavor mixing indeed exists~\cite{Xing:2011zza, Zyla:2020}. In order to generate tiny neutrino masses, one can extend the standard model (SM) by adding a few new particles and allowing for the lepton number violation, and then the tiny masses of 	light neutrinos can be attributed to the introduced heavy degrees of freedom. This is the so-called seesaw mechanism. For example, in the typical type-I seesaw mechanism~\cite{Minkowski:1977sc, Yanagida:1979, GellMan1979, Mohapatra:1979ia}, three right-handed neutrinos, which are singlets under the ${\rm SU}(2)_{\rm L} \times {\rm U}(1)^{}_{\rm Y}$ gauge symmetry of the SM, are introduced and the smallness of light neutrino masses can thus be explained by the heavy mass scale of the right-handed neutrinos.

Another interesting realization of the seesaw mechanism is the type-II seesaw mechanism~\cite{Konetschny:1977bn,Magg:1980ut,Schechter:1980gr,Cheng:1980qt,Lazarides:1980nt,Mohapatra:1980yp}, in which an additional Higgs triplet under ${\rm SU(2)^{}_{\rm L}}$ is added into the SM. Therefore, the gauge-invariant Lagrangian relevant for lepton masses and flavor mixing can be written as
\begin{eqnarray}\label{eq:LagSM}
-{\cal L}^{}_{\rm lepton} = \overline{\ell^{}_{\rm L}} Y^{}_l H E^{}_{\rm R} +  \frac{1}{2} \overline{\ell^{}_{\rm L}} Y^{}_{\Delta}    \Delta {\rm i} \sigma^{}_{2}\ell^{\rm c}_{\rm L} 
+ {\rm h.c.} \; ,
\end{eqnarray}
where $\ell^{}_{\rm L}$ and $E^{}_{\rm R}$ denote the left-handed lepton doublet and the right-handed charged-lepton singlet, $H$ and $\Delta$ are the Higgs doublet and triplet, respectively. Note that in Eq.~(\ref{eq:LagSM}), $\ell^{\rm c}_{\rm L} \equiv C\overline{\ell^{}_{\rm L}}^{\rm T}_{}$ with $C={\rm i}\gamma^{2}_{}\gamma^{0}_{}$ being the charge-conjugation matrix has been defined. After the spontaneous symmetry breaking, we can obtain the charged-lepton and neutrino mass matrices as $M^{}_{l}=Y^{}_{l}v/\sqrt{2}$ and $M^{}_{\nu}=Y^{}_{\Delta}v^{}_{\Delta}$ respectively, where $v =\sqrt{2}\langle H^{0}_{}\rangle \approx 246~{\rm GeV}$ with $\langle H^{0}_{}\rangle$ being the vacuum expectation value (vev) of the neutral component of $H$ and $v^{}_{\Delta}$ is the vev of the neutral component of $\Delta$. 
The explicit form of $v^{}_{\Delta}$ can be determined from the following potential which involves both the Higgs doublet and triplet
\begin{eqnarray}
V(H,\Delta)=-\mu^2_{}H^{\dag}_{}H+\lambda(H^{\dag}_{}H)^2_{}+\dfrac{1}{2}M^2_{\Delta}{\rm Tr}(\Delta^{\dag}_{}\Delta)-(\lambda^{}_\Delta M^{}_\Delta H^{\rm T}_{} {\rm i}\sigma^{}_2 \Delta H +{\rm h.c.}) \; ,
\label{eq:potenHD}
\end{eqnarray}
where $\mu$, $\lambda$ and $\lambda^{}_\Delta$ are the coupling coefficients and $M^{}_{\Delta}$ denotes the mass of the Higgs triplet. Then the vev's $v=\sqrt{\mu^2/(\lambda-2\lambda^{2}_{\Delta})}$ and $v^{}_\Delta=\lambda^{}_\Delta v^2_{}/M^{}_\Delta$ can be derived from Eq.~(\ref{eq:potenHD}). The small value of $v^{}_\Delta$, which is suppressed by the large mass scale of $M^{}_\Delta$, can also explain the observed tiny neutrino masses. In the framework of SO(10) grand unified theories~\cite{Fritzsch:1974nn}, both right-handed neutrinos and the Higgs triplet are naturally embedded. Therefore the light neutrino masses receive the contributions from both right-handed neutrinos and the Higgs triplet via the type-(I+II) seesaw mechanism. In this case, the Lagrangian in Eq.~(\ref{eq:LagSM}) becomes
\begin{eqnarray}\label{eq:LagSM1}
-{\cal L}^{}_{\rm lepton} = \overline{\ell^{}_{\rm L}} Y^{}_l H E^{}_{\rm R} + \overline{\ell^{}_{\rm L}} Y^{}_{\nu} \widetilde{H} N^{}_{\rm R} +\frac{1}{2}\overline{N^{\rm c}_{\rm R}} M^{}_{\rm R}N^{}_{\rm R} + \frac{1}{2} \overline{\ell^{}_{\rm L}} Y^{}_{\Delta}  \Delta {\rm i} \sigma^{}_{2} \ell^{\rm c}_{\rm L} 
+ {\rm h.c.} \; ,
\end{eqnarray}
where $N^{}_{\rm R}$ and $M^{}_{\rm R}$ denote the right-handed neutrinos and their Majorana mass matrix respectively. Note that in Eq.~(\ref{eq:LagSM1}), $\widetilde{H} \equiv {\rm i}\sigma^{}_{2}H^{\ast}_{}$ and $N^{\rm c}_{\rm R} \equiv C\overline{N^{}_{\rm R}}^{\rm T}_{}$ have been defined.

Although the seesaw model provides us with an elegant way to explain the tiny neutrino masses, it can not account for the flavor structures existing in the lepton mass matrices. As a consequence, the model is in general lacking of predictive power for lepton mass spectra, flavor mixing pattern and CP violation~\cite{Xing:2019vks}. On this account, non-Abelian discrete flavor symmetries have been implemented in the seesaw model to explain the flavor mixing in recent literature, e.g., Refs.~\cite{Altarelli:2010gt, Ishimori:2010au, King:2013eh, King:2014nza, King:2017guk,Feruglio:2019ktm}. To be specific, one can first assume that the Lagrangian maintains an overall discrete flavor symmetry at some high-energy scale. Next a few of gauge-singlet scalar fields which are called flavons are introduced to break down the whole symmetry into distinct residual symmetries in the charged-lepton and neutrino sectors~\cite{Lam:2008rs, Lam:2008sh, Ge:2011ih, Ge:2011qn, Hernandez:2012ra, Feruglio:2012cw}. Then the flavor structures will be determined by the vev's of these flavons. However, the introduction of flavons will inevitably bring a large number of free parameters into the model and how to experimentally prove the existence of the flavons is also a tough problem. 

Recently, a new and attractive approach to solve the flavor mixing problem, which is based on the modular invariance, has been proposed in Ref.~\cite{Feruglio:2017spp}. Within the framework of modular symmetries, the Yukawa couplings are regarded as the modular forms with even weights, which transform as the multiplets under some finite modular symmetry groups $\Gamma^{}_N$. For a given value of $N$, $\Gamma^{}_N$ is isomorphic to the well-known non-Abelian discrete symmetry group, e.g., $\Gamma^{}_{2} \simeq S^{}_{3}$~\cite{Kobayashi:2018vbk, Kobayashi:2018wkl, Kobayashi:2019rzp, Okada:2019xqk}, $\Gamma^{}_{3} \simeq A^{}_{4}$~\cite{Kobayashi:2018scp, Criado:2018thu, deAnda:2018ecu, Okada:2018yrn, Nomura:2019jxj, Nomura:2019lnr, Ding:2019zxk, Nomura:2019yft, Okada:2019mjf, Asaka:2019vev}, $\Gamma^{}_{4} \simeq S^{}_4$~\cite{Penedo:2018nmg, Novichkov:2018ovf, Okada:2019lzv,Wang:2019ovr} and $\Gamma^{}_{5} \simeq A^{}_5$~\cite{Novichkov:2018nkm, Ding:2019xna, Criado:2019tzk}. Modular forms are the functions of the modulus $\tau$. Once the value of $\tau$ is given, the modular symmetry is always broken and then the flavor mixing pattern can be generated. Therefore the flavon field is not necessary in the minimal framework of modular symmetries. Except the references we have mentioned above, there are also plenty of works related to other interesting aspects of modular symmetries, such as the combination of modular symmetries and the CP symmetry~\cite{Novichkov:2019sqv,Kobayashi:2019uyt}, multiple modular symmetries~\cite{deMedeirosVarzielas:2019cyj, King:2019vhv}, the double covering of modular groups~\cite{Liu:2019khw}, the $A^{}_{4}$ symmetry from the modular $S^{}_{4}$ symmetry~\cite{Kobayashi:2019mna, Kobayashi:2019xvz}, the modular residual symmetry~\cite{Novichkov:2018yse,Gui-JunDing:2019wap}, the unification of quark and lepton flavors with modular invariance~\cite{Okada:2019uoy,Okada:2020rjb}, the realization of texture zeros via the modular symmetry~\cite{Zhang:2019,Lu:2019vgm} and the applications of modular symmetries on other types of seesaw models~\cite{Kobayashi:2019gtp,Nomura:2019xsb}.

In this paper, we investigate the minimal supersymmetric type-(I+II) seesaw model~\cite{Gu:2006wj}, where only one right-handed neutrino and two Higgs triplets are introduced, with the modular $A^{}_4$ symmetry and explore its implications for lepton mass spectra, flavor mixing pattern and CP violation. Such an investigation has strong theoretical motivations. First, the minimal type-(I+II) seesaw model itself is a very economical model with only a few degrees of freedom, so it is meaningful to study the implications of the modular symmetry in constructing a minimal type-(I+II) seesaw model. Second, we notice that different from the type-I seesaw mechanism, the type-II seesaw model with the modular symmetry is only discussed in Ref.~\cite{Kobayashi:2019gtp}. And as can be seen in that paper, one has to require higher weights of the modular forms and a large number of free model parameters in the modular type-II seesaw model to obtain feasible parameter space, which makes the model less predictive. While if we consider the minimal type-(I+II) seesaw model, i.e., an additional right-handed neutrino is introduced, the number of free parameters can be reduced to eight. Finally, the charge assignments of superfields and modular forms can be tightly restricted due to the existence of both the right-handed neutrino and Higgs triplets in the minimal type-(I+II) seesaw model. We present a detailed analysis of the low-energy phenomenology of our model and show that it can be consistent with current experimental data within $1\sigma$ level. Our model predicts relatively large masses of three light neutrinos, and the individual contributions to the neutrino masses from the right-handed neutrino and the Higgs triplet turn out to be comparable. The octant of the mixing angle $\theta^{}_{23}$ strongly depends on the hierarchy of free parameters in the charged-lepton sector. Furthermore, the values of $\theta^{}_{23}$ and Dirac CP-violating phase $\delta$ can reach $45^{\circ}_{}$ and $90^{\circ}_{}$ (or $270^{\circ}_{}$) respectively,  indicating the neutrino mass matrix possesses an approximate $\mu-\tau$ reflection symmetry~\cite{Harrison:2002et} for some specific values of free parameters. 

The remaining part of this paper is organized as follows. In Sec.~\ref{sec:modular}, a brief summary of the modular $A^{}_4$ symmetry is given. The concrete type-(I+II) seesaw model with the modular $A^{}_4$ symmetry is then proposed in Sec.~\ref{sec:models}. The low-energy phenomenology of lepton mass spectra, flavor mixing pattern and CP violation in our model are discussed in Sec.~\ref{sec:lowphe}. Finally, we summarize our main conclusions in Sec.~\ref{sec:summary}. Some properties of the modular $A^{}_4$ symmetry group are presented in Appendix~\ref{sec:appA}.

\section{Modular $A^{}_4$ Symmetry} \label{sec:modular}
The basics of modular symmetries have been expounded in previous works (See, e.g., Ref.~\cite{Feruglio:2017spp}). In this section, we shall only give a brief review on the modular symmetry. 

In a supersymmetric theory, the action ${\cal S}$ keeps invariant under the modular transformations
\begin{eqnarray}
\gamma: \tau \rightarrow \dfrac{a \tau + b}{c \tau + d} \; , \quad
\chi^{(I)}_{} \rightarrow (c \tau +d )^{-k^{}_I} \rho^{(I)}_{} (\gamma) \chi^{(I)} _{} \; ,
\label{eq:actiontransf}
\end{eqnarray}
where $\gamma$ is the element of the modular group $\Gamma$ with $a$, $b$, $c$ and $d$ being integers satisfying $ad - bc =1$, $\tau$ is an arbitrary complex number in the upper complex plane, $\rho^{(I)}(\gamma)$ denotes the representation matrix of the modular transformation $\gamma$, and $k^{}_{I}$ is the weight associated with the supermultiplet $\chi^{(I)}$. As a consequence, the K\"ahler potential ${\cal K}(\tau,\overline{\tau}, \chi, \overline{\chi})$ is invariant up to the K\"ahler transformation ${\cal K}(\tau,\overline{\tau}, \chi, \overline{\chi}) \rightarrow {\cal K}(\tau,\overline{\tau}, \chi, \overline{\chi}) + f(\tau, \chi)+f(\overline{\tau}, \overline{\chi})$\footnote{The most general K\"ahler potential consistent with the modular symmetry may contain additional terms, as recently pointed out in Ref.~\cite{Chen:2019ewa}. However, for a phenomenological purpose, we consider only the simplest form of the K\"ahler potential.}, where $f(\tau, \chi)$ is a holomorphic function, which is not necessarily modular invariant. Meanwhile, the superpotential ${\cal W}(\tau, \chi)$ is invariant as well and can be expanded in terms of the supermultiplets as follows
\begin{eqnarray}
{\cal W}(\tau, \chi)= \sum_{n}^{}\sum_{\{I^{}_{1},\dots,I^{}_{n}\}}^{} Y^{}_{I^{}_1\dots I^{}_n}(\tau)\chi_{}^{(I^{}_1)}\cdots\chi_{}^{(I^{}_n)} \; ,
\label{eq:surpoten}
\end{eqnarray}
where the coefficients $Y^{}_{I^{}_1\dots I^{}_n}(\tau)$ take the modular forms, transforming under the finite modular group $\Gamma^{}_{N} \equiv \Gamma/\Gamma(N)$ (with $\Gamma(N)$ being the principal congruence subgroup of $\Gamma$) as
\begin{eqnarray}
Y^{}_{I^{}_1\dots I^{}_n}(\tau) \rightarrow (c\tau+d)^{k^{}_Y}_{} \rho^{}_{Y} (\gamma) Y^{}_{I^{}_1 \dots I^{}_n}(\tau) \; ,
\label{eq:Yuktransf}
\end{eqnarray}
where the even integer $k^{}_{Y}$ is the weight of $Y^{}_{I^{}_1\dots I^{}_n}(\tau)$ and $\rho^{}_Y$ is the representation matrix of $\Gamma^{}_N$. In addition, $k^{}_Y$ and $\rho^{}_{Y}$ must satisfy $k^{}_{Y} = k^{}_{I^{}_1} +  \cdots + k^{}_{I^{}_N}$ and $\rho^{}_{Y} \otimes \rho^{(I^{}_{1})}_{}  \otimes \cdots \otimes \rho^{(I^{}_{N})}_{} \ni {\bf 1}$, respectively.

For the symmetry group $\Gamma^{}_{3} \simeq A^{}_{4}$ of our interest, there are three linearly independent modular forms of the lowest non-trivial weight $k^{}_{Y}=2$, denoted as $Y^{}_i(\tau)$ for $i = 1, 2, 3$, which form a triplet ${\bf 3}$ under the modular $A^{}_4$ symmetry transformations~\cite{Feruglio:2017spp}, namely,
\begin{eqnarray}
Y^{}_{\bf 3} (\tau) \equiv  \left(\begin{matrix} Y^{}_{1}(\tau) \\ Y^{}_2 (\tau) \\ Y^{}_{3} (\tau) \end{matrix}\right) \; .
\label{eq:A4Y}
\end{eqnarray}
The exact expressions of $Y^{}_{i}(\tau)$ (for $i=1,2,3$) are presented in Appendix~\ref{sec:appA}. Based on the modular forms $Y^{}_{i}(\tau)$ of weight $k^{}_Y = 2$, one can construct the modular forms of higher weights, such as $k^{}_Y = 4$ and $k^{}_Y = 6$. For $k^{}_Y = 4$, there are totally five independent modular forms, which transform as {\bf 1}, ${\bf 1}^{\prime}_{}$ and {\bf 3} under the $A^{}_{4}$ symmetry~\cite{Feruglio:2017spp,Ding:2019zxk,Zhang:2019}, namely,
\begin{equation}
Y^{(4)}_{\bf 1} = Y^{2}_{1}+2Y^{}_{2}Y^{}_{3} \; , \quad Y^{(4)}_{{\bf 1}^{\prime}_{}} = Y^2_{3}+2Y^{}_1Y^{}_2 \; , \quad 
Y^{(4)}_{\bf 3} =\left( \begin{matrix} Y^2_1-Y^{}_2Y^{}_3 \\ Y^2_3-Y^{}_1Y^{}_2 \\ Y^2_2-Y^{}_1Y^{}_3 \end{matrix} \right) \; ,
\label{eq:Y4}
\end{equation}
where the argument $\tau$ of all the modular forms is suppressed. For $k^{}_Y = 6$, we have seven independent modular forms, whose assignments under the $A^{}_4$ symmetry are as follows~\cite{Feruglio:2017spp,Ding:2019zxk,Zhang:2019}
\begin{eqnarray}
Y^{(6)}_{\bf 1} &=& Y^3_1+Y^3_2+Y^3_3-3Y^{}_1Y^{}_2Y^{}_3 \; , \nonumber \\
Y^{(6)}_{{\bf 3},1} &=& (Y^2_1+2Y^{}_2Y^{}_3)\left(
\begin{matrix} 
Y^{}_{1} \\ Y^{}_{2} \\ Y^{}_{3} 
\end{matrix} \right) \; , 
\quad Y^{(6)}_{{\bf 3},2} = (Y^2_3+2Y^{}_1Y^{}_2)\left(
\begin{matrix} 
Y^{}_{3} \\ Y^{}_{1} \\ Y^{}_{2} 
\end{matrix} \right) \; .
\label{eq:Y6}
\end{eqnarray}

\section{The Minimal Type-(I+II) Seesaw Model} \label{sec:models}
In this section, we are going to construct a minimal type-(I+II) seesaw model with the modular $A^{}_{4}$ symmetry. To begin with, let us first make some general remarks on the model building.
\begin{itemize}
	\item  A criterion for the model building is that our model should be economical enough, which means that the number of free model parameters should be as small as possible. To be specific, we have eight low-energy observables, including three charged-lepton masses $\{m^{}_e, m^{}_\mu, m^{}_\tau\}$, two independent neutrino mass-squared differences $\{\Delta m^2_{21}, \Delta m^2_{31}\}$ in the normal mass  ordering (NO) case where $m^{}_1<m^{}_2<m^{}_3$ or $\{\Delta m^2_{21}, \Delta m^2_{32}\}$ in the inverted mass ordering (IO) case where $m^{}_3<m^{}_1<m^{}_2$ and three mixing angles $\{\theta^{}_{12}, \theta^{}_{13}, \theta^{}_{23}\}$. Therefore the number of free model parameters should be no more than eight in order to have predictive power for the other parameters, such as the CP-violating phases.
	\item  As the modular symmetry is intrinsically working in the supersymmetric framework, we should introduce one chiral superfield $\widehat{N}^{\rm c}_{}$ \footnote {In this paper, we use the ``hat'' symbol to denote the chiral superfield.} which contains the right-handed neutrino singlet and a pair of ${\rm SU}(2)^{}_{\rm L}$ triplet Higgs superfields $\{\widehat{\Delta}^{}_{\rm 1},\widehat{\Delta}^{}_{2}\}$ with the hypercharges $\{+1,-1\}$ defined as
	\begin{eqnarray}
	\widehat{\Delta}^{}_{1} \equiv \sqrt{2} \left(
	\begin{matrix}
	\widehat{\Delta}^{+}_{\rm 1}/\sqrt{2} && \widehat{\Delta}^{++}_{\rm 1} \\ \widehat{\Delta}^{0}_{\rm 1} && -\widehat{\Delta}^{+}_{\rm 1}/\sqrt{2}
	\end{matrix}\right) \; , \quad 
	\widehat{\Delta}^{}_{2} \equiv \sqrt{2}\left(
	\begin{matrix}
	\widehat{\Delta}^{-}_{\rm 2}/\sqrt{2} && \widehat{\Delta}^{0}_{\rm 2} \\ \widehat{\Delta}^{--}_{\rm 2} && -\widehat{\Delta}^{-}_{\rm 2}/\sqrt{2}
	\end{matrix}\right) \; ,
	\label{eq:Higgstri}
	\end{eqnarray} 
	where $\widehat{\Delta}^{++}_{\rm 1}$ ($\widehat{\Delta}^{--}_{\rm 2}$), $\widehat{\Delta}^{+}_{\rm 1}$ ($\widehat{\Delta}^{-}_{\rm 2}$) and $\widehat{\Delta}^{0}_{\rm 1}$ ($\widehat{\Delta}^{0}_{\rm 2}$) denote the doubly-charged, singly-charged and neutral components of $\widehat{\Delta}^{}_{1}$ ($\widehat{\Delta}^{}_{2}$), respectively. $\widehat{N}^{\rm c}$, $\widehat{\Delta}^{}_{1}$ and $\widehat{\Delta}^{}_{2}$ are all arranged to be the trivial singlet under the modular $A^{}_{4}$ symmetry in our model for simplicity. Furthermore, the superfields for Higgs doublets $\{\widehat{H}^{}_{\rm u}, \widehat{H}^{}_{\rm d}\}$ with the hypercharges $\{+1/2,-1/2\}$ are also assigned into {\bf 1} under the modular $A^{}_{4}$ symmetry. As a consequence, we do not need to change the remaining part of the MSSM irrelevant for leptonic flavor mixing.
	\item The superfields for three lepton doublets $\{\widehat{L}^{}_1, \widehat{L}^{}_2, \widehat{L}^{}_3\}$ are arranged as a triplet ${\bf 3}$ under the $A^{}_4$ symmetry, while the superfields for three charged-lepton singlets $\{\widehat{E}^{\rm c}_1, \widehat{E}^{\rm c}_2, \widehat{E}^{\rm c}_3\}$ should be assigned into three different singlets of $A^{}_4$ (e.g., $\widehat{E}^{\rm c}_{1} \sim {\bf 1}$, $\widehat{E}^{\rm c}_{2} \sim {\bf 1}^{\prime\prime}$ and $\widehat{E}^{\rm c}_{3} \sim {\bf 1}^{\prime}$). Otherwise, it will be difficult to explain the strong mass hierarchy of three charged leptons, namely, $m^{}_e \ll m^{}_\mu \ll m^{}_\tau$.
	\item The modular forms relevant for lepton masses and flavor mixing can be exactly determined from the two identities $k^{}_{Y} = k^{}_{I^{}_1} +  \cdots + k^{}_{I^{}_N}$ and $\rho^{}_{Y} \otimes \rho^{(I^{}_{1})}_{}  \otimes \cdots \otimes \rho^{(I^{}_{N})}_{} \ni {\bf 1}$ after the weights and representations of the superfields are fixed. Note that since both the right-handed neutrino and Higgs triplets are introduced into our model, more terms will appear in the whole superpotential. Consequently, there remains less freedom for us to adjust the weights and representations under the modular $A^{}_4$ symmetry of all the superfields as well as the modular forms. In Table~\ref{Table:Assign}, we show the charge assignments of the chiral superfields and the couplings under the ${\rm SU}(2)^{}_{\rm L}$ gauge symmetry and the modular $A^{}_{4}$ symmetry for our model, and the corresponding modular weights are listed in the last row. Note that $k^{}_{\rm D}=4$ and $k^{}_{\rm R}=6$ are the lowest weights which the modular forms $f^{}_{\rm D}$ and $f^{}_{\rm R}$ can take respectively under the premise that $k^{}_{Y} = k^{}_{I^{}_1} +  \cdots + k^{}_{I^{}_N}$ should be satisfied in each superpotential.	
\end{itemize}

\begin{table}[t!]
	\centering
		\vspace{-0.25cm}\caption{The charge assignment of the chiral superfields and the relevant couplings under the ${\rm SU(2)^{}_{\rm L}}$ symmetry and the modular $A^{}_{4}$ symmetry in our model, with the corresponding modular weights listed in the last row.}\vspace{0.5cm}
	\begin{tabular}{ccccccccc}
		\toprule 
		& $\widehat{L}$ & $\widehat{E}^{\rm c}_{1}, \widehat{E}^{\rm c}_{2}, \widehat{E}^{\rm c}_{3}$ & $\widehat{N}^{\rm c}_{}$  & $\widehat{H}^{}_{\rm u},\widehat{H}^{}_{\rm d}$ &  $\widehat{\Delta}^{}_{1},\widehat{\Delta}^{}_{2}$  & $f^{}_{e}(\tau), f^{}_{\mu}(\tau), f^{}_{\tau}(\tau), f_{\Delta}(\tau)$ & $f^{}_{\rm D}(\tau)$ & $f^{}_{\rm R}(\tau)$\\
		\midrule
		{\rm SU(2)} & 2 & 1 & 1  & 1 & 3 & 1 & 1 & 1\\
		$A^{}_{4}$ & \bf{3} & $\bf{1},\bf{1^{\prime\prime}_{}},\bf{1^{\prime}_{}} $& \bf{1}  & \bf{1} & \bf{1} & \bf{3} & \bf{3} &\bf{1} \\
		$-k^{}_{I}$ & $-1$ & $-1$ & $-3$ & 0 & 0 & $k^{}_{e,\mu,\tau,\Delta}=2$ & $k^{}_{\rm D}=4$ & $k^{}_{\rm R}=6$\\
		\bottomrule
	\end{tabular}
	\label{Table:Assign}
\end{table} 
Keeping these assignments above in mind, now it is straightforward for us to write down the modular $A^{}_{4}$ invariant superpotential ${\cal W}$, which can be decomposed into three parts ${\cal W} = {\cal W}^{}_{l} + {\cal W}^{}_{\rm I} + {\cal W}^{}_{\rm II}$ with
\begin{eqnarray}
{\cal W}^{}_{l} &=& \alpha^{}_{1} \left[\left(\widehat{L} \widehat{E}^{\rm c}_{1}\right)^{}_{\bf 3}(f^{}_{e}(\tau))^{}_{\bf 3}\right]^{}_{\bf 1}\widehat{H}^{}_{\rm d}+\alpha^{}_{2}\left[\left(\widehat{L} \widehat{E}^{\rm c}_{2}\right)^{}_{\bf 3}(f^{}_{\mu}(\tau))^{}_{\bf 3}\right]^{}_{\bf 1}\widehat{H}^{}_{\rm d}+\alpha^{}_{3}\left[\left(\widehat{L} \widehat{E}^{\rm c}_{3}\right)^{}_{\bf 3}(f^{}_{\tau}(\tau))^{}_{\bf 3}\right]^{}_{\bf 1}\widehat{H}^{}_{\rm d} \; , \nonumber \\
{\cal W}^{}_{\rm I} &=& g^{}_{1} \left[\left(\widehat{L} \widehat{N}^{\rm c}_{}\right)^{}_{\bf 3} (f^{}_{\rm D}(\tau))^{}_{\bf 3}\right]^{}_{\bf 1}\widehat{H}^{}_{\rm u}+\frac{1}{2} \Lambda \left[ \left(\widehat{N}^{\rm c}_{}\widehat{N}^{\rm c}_{}\right)^{}_{\bf 1} (f^{}_{\rm R}(\tau))^{}_{\bf 1}\right]^{}_{\bf 1}  \; , \nonumber \\
{\cal W}^{}_{\rm II} &=& \frac{1}{2}g^{}_{2} \left[\left(\widehat{L}\widehat{L}\right)^{}_{\bf 3}(f^{}_{\Delta})^{}_{\bf 3}\right]^{}_{\bf 1}\left({\rm i}\sigma^{}_{2} \widehat{\Delta}^{}_{1}\right)\; , \label{eq:superp1} 
\end{eqnarray}
where $\alpha^{}_{1}$, $\alpha^{}_{2}$ and $\alpha^{}_{3}$ are three coupling coefficients in the charged-lepton sector which we can set to be real and positive without loss of generality while $g^{}_{1}$, $g^{}_{2}$ and $\Lambda$ are the coupling coefficients in the neutrino sector. In Eq.~(\ref{eq:superp1}), the individual contributions to neutrino masses from the type-I and type-II seesaw mechanisms can be read from ${\cal W}^{}_{\rm I}$ and ${\cal W}^{}_{\rm II}$, respectively. When the modulus parameter $\tau$ is fixed, the modular symmetry is broken down and the superpotential reads
\begin{eqnarray}
{\cal W} = \widehat{L}^{\rm T}_{} \lambda^{}_{l} \widehat{E}^{\rm c}_{} \widehat{H}^{}_{{\rm d}} + \widehat{L}^{\rm T}_{} \lambda^{}_{\rm D} \widehat{N}^{\rm c} \widehat{H}^{}_{\rm u}  + \frac{1}{2} (\widehat{N}^{\rm c}_{})^{\rm T}_{} \lambda^{}_{\rm R} \widehat{N}^{\rm c}_{} + \frac{1}{2} \widehat{L}^{\rm T}_{} \lambda^{}_{\rm II}   \widehat{L} \left({\rm i}\sigma^{}_{2} \widehat{\Delta}^{}_{1} \right)  \; ,
\label{eq:superp2}
\end{eqnarray}
where $\lambda^{}_{l}$ and $\lambda^{}_{\rm D}$ turn out to be the charged-lepton and Dirac neutrino Yukawa coupling matrices, respectively, $\lambda^{}_{\rm R}$ becomes the right-handed neutrino mass matrix and $\lambda^{}_{\rm II}$ is the neutrino mass matrix induced by the type-II seesaw. 

On the other hand, the superpotential relevant for the couplings between the Higgs doublets and triplets, which is just the supersymmetric version of Eq.~(\ref{eq:potenHD}), can be written as~\cite{Rossi:2002zb}
\begin{eqnarray}
{\cal W}^{}_{\rm H-\Delta} = \mu \widehat{H}^{}_{\rm u}\widehat{H}^{}_{\rm d} +\lambda^{}_{1}\widehat{H}^{}_{\rm d}\left({\rm i}\sigma^{}_{2}\widehat{\Delta}^{}_{1}\right)\widehat{H}^{}_{\rm d}+\lambda^{}_{2}\widehat{H}^{}_{\rm u}\left({\rm i}\sigma^{}_{2}\widehat{\Delta}^{}_{2}\right)\widehat{H}^{}_{\rm u}+ \frac{1}{2}M^{}_{\Delta}{\rm Tr}\left(\widehat{\Delta}^{}_{1}\widehat{\Delta}^{}_{2}\right) \; ,
\label{eq:superp3}
\end{eqnarray}
where $\mu$, $\lambda^{}_{1}$ and $\lambda^{}_{2}$ are the coupling coefficients. After the supersymmetry breaking and the ${\rm SU(2)}^{}_{\rm L}\times{\rm U(1)}^{}_{\rm Y}$ gauge symmetry breakdown, all the Higgs fields get their own vev's and one can then obtain the lepton mass terms from Eq.~(\ref{eq:superp1}). It has been indicated in Ref.~\cite{Wang:2019ovr} that there exists the following correspondence between the lepton mass matrices and the Yukawa coupling matrices in the MSSM framework under the left-right convention for the fermion mass terms
\begin{eqnarray}
M^{}_l = v^{}_{\rm d} \lambda^{\ast}_l/\sqrt{2} \; , \quad  M^{}_{\rm D} = v^{}_{\rm u} \lambda^{\ast}_{\rm D}/\sqrt{2} \; ,  \quad  M^{}_{\rm R} = \lambda^{\ast}_{\rm R} \; , \quad  M^{}_{\rm II} = v^{}_{1}\lambda^{\ast}_{\rm II} \; , \label{eq:Mlambda}
\end{eqnarray} 
where $v^{}_{\rm d} = v \cos\beta$ and $v^{}_{\rm u} = v\sin\beta$ are respectively the vev of the neutral scalar component field of $\widehat{H}^{}_{\rm d}$ to that of $\widehat{H}^{}_{\rm u}$, with $\tan\beta \equiv v^{}_{\rm u}/v^{}_{\rm d}$ being their ratio, and $v^{}_{1}=\lambda^{}_{2}v^{2}_{\rm u}/M^{}_{\Delta}$ is the vev of the neutral scalar component field of $\widehat{\Delta}^{}_{1}$ and can be derived from Eq.~(\ref{eq:superp3}). Note that here we use``$\ast$'' to denote the complex conjugation. Therefore, by using the product rules of the $A^{}_{4}$ symmetry group collected in Appendix~\ref{sec:appA}, we can obtain the charged-lepton mass matrix
\begin{eqnarray}
M^{}_{l} = \frac{v^{}_{\rm d}}{\sqrt{2}}\left(
\begin{matrix}
Y^{}_{1} && Y^{}_{2} && Y^{}_{3} \\
Y^{}_{3} && Y^{}_{1} && Y^{}_{2} \\
Y^{}_{2} && Y^{}_{3} && Y^{}_{1} 
\end{matrix}
\right)^{\ast}_{}
\left(
\begin{matrix}
\alpha^{}_{1} && 0 && 0 \\
0 && \alpha^{}_{2} && 0  \\
0 && 0 && \alpha^{}_{3} 
\end{matrix}
\right) \; ,
\label{eq:Me}
\end{eqnarray}
and the Dirac neutrino mass matrix
\begin{eqnarray}
M^{}_{\rm D} =\frac{v^{}_{\rm u}g^{\ast}_{1}}{\sqrt{2}}\left(
\begin{matrix}
Y^{2}_{1}-Y^{}_{2}Y^{}_{3} \\
Y^{2}_{2}-Y^{}_{1}Y^{}_{3} \\
Y^{2}_{3}-Y^{}_{1}Y^{}_{2}
\end{matrix}\right)^{\ast}_{}\; .
\label{eq:MD}
\end{eqnarray}
Since only one right-handed neutrino is introduced, the Majorana mass matrix will degenerate to a complex number
\begin{eqnarray}
M^{}_{\rm R} = \Lambda(Y^{3}_{1}+Y^{3}_{2}+Y^{3}_{3}-3Y^{}_{1}Y^{}_{2}Y^{}_{3})^{\ast}_{} \; .
\label{eq:MR}
\end{eqnarray}  
After applying the type-I seesaw mechanism $M^{}_{\rm I} \approx -M^{}_{\rm D}M^{-1}_{\rm R}M^{\rm T}_{\rm D}$, we arrive at the neutrino mass matrix from the type-I seesaw
\begin{eqnarray}
M^{}_{\rm I} = &&-\frac{v^{2}_{\rm u}(g^{\ast}_{1})^{2}_{}}{2\Lambda(Y^{3}_{1}+Y^{3}_{2}+Y^{3}_{3}-3Y^{}_{1}Y^{}_{2}Y^{}_{3})^{\ast}_{}}
\nonumber \\
&&\times \left(\begin{matrix}
(Y^{2}_{1}-Y^{}_{2}Y^{}_{3})^{2}_{} && (Y^{2}_{1}-Y^{}_{2}Y^{}_{3})(Y^{2}_{2}-Y^{}_{1}Y^{}_{3}) && (Y^{2}_{1}-Y^{}_{2}Y^{}_{3})(Y^{2}_{3}-Y^{}_{1}Y^{}_{2}) \\
(Y^{2}_{1}-Y^{}_{2}Y^{}_{3})(Y^{2}_{2}-Y^{}_{1}Y^{}_{3}) && (Y^{2}_{2}-Y^{}_{1}Y^{}_{3})^2_{} && (Y^{2}_{2}-Y^{}_{1}Y^{}_{3})(Y^{2}_{3}-Y^{}_{1}Y^{}_{2}) \\
(Y^{2}_{1}-Y^{}_{2}Y^{}_{3})(Y^{2}_{3}-Y^{}_{1}Y^{}_{2})  && (Y^{2}_{2}-Y^{}_{1}Y^{}_{3})(Y^{2}_{3}-Y^{}_{1}Y^{}_{2}) && (Y^{2}_{3}-Y^{}_{1}Y^{}_{2})^2_{}
\end{matrix}\right)^{\ast}_{} \;. \nonumber \\
\label{eq:Mnu1}
\end{eqnarray}
Meanwhile, the neutrino mass matrix induced by the type-II seesaw can be expressed as
\begin{eqnarray}
M^{}_{\rm II} =  
\frac{\lambda^{}_{2}v^{2}_{\rm u}g^{\ast}_{2}}{3M^{}_{\Delta}}\left(\begin{matrix}
2Y^{}_{1} && -Y^{}_{3} && -Y^{}_{2} \\
-Y^{}_{3} && 2Y^{}_{2} && -Y^{}_{1} \\
-Y^{}_{2} && -Y^{}_{1} && 2Y^{}_{3}
\end{matrix}\right)^{\ast}_{} \; .
\label{eq:Mnu2}
\end{eqnarray}
Then the whole effective neutrino mass matrix should be a combination of $M^{}_{\rm I}$ and $M^{}_{\rm II}$. Since the overall phase of any lepton mass matrix is irrelevant for lepton masses and flavor mixing, one can take $g^{}_1$ in Eq.~(\ref{eq:MD}) to be real and it is convenient to define a new complex parameter as $  2\lambda^{}_{\rm 2} g^{}_{2} \Lambda /(3 g^{2}_{1} M^{}_{\Delta}) \equiv \widetilde{g} =g e^{{\rm i}\phi^{}_{g}}$ with $g = |\widetilde{g}|$ and $\phi^{}_g \equiv \arg(\widetilde{g})$. Therefore $M^{}_{\nu}$ can be written as
\begin{eqnarray}
M^{}_{\nu}= && M^{}_{\rm I}+M^{}_{\rm II} \nonumber \\
=&&-\frac{v^{2}_{\rm u} g^{2}_{1}}{2\Lambda}\left[\rule{0cm}{1cm}\right.\frac{1}{Y^{3}_{1}+Y^{3}_{2}+Y^{3}_{3}-3Y^{}_{1}Y^{}_{2}Y^{}_{3}}
\nonumber \\
&&\times \left(\begin{matrix}
(Y^{2}_{1}-Y^{}_{2}Y^{}_{3})^{2}_{} && (Y^{2}_{1}-Y^{}_{2}Y^{}_{3})(Y^{2}_{2}-Y^{}_{1}Y^{}_{3}) && (Y^{2}_{1}-Y^{}_{2}Y^{}_{3})(Y^{2}_{3}-Y^{}_{1}Y^{}_{2}) \\
(Y^{2}_{1}-Y^{}_{2}Y^{}_{3})(Y^{2}_{2}-Y^{}_{1}Y^{}_{3}) && (Y^{2}_{2}-Y^{}_{1}Y^{}_{3})^2_{} && (Y^{2}_{2}-Y^{}_{1}Y^{}_{3})(Y^{2}_{3}-Y^{}_{1}Y^{}_{2}) \\
(Y^{2}_{1}-Y^{}_{2}Y^{}_{3})(Y^{2}_{3}-Y^{}_{1}Y^{}_{2})  && (Y^{2}_{2}-Y^{}_{1}Y^{}_{3})(Y^{2}_{3}-Y^{}_{1}Y^{}_{2}) && (Y^{2}_{3}-Y^{}_{1}Y^{}_{2})^2_{}
\end{matrix}\right) \nonumber\\
&& -\widetilde{g} \left(\begin{matrix}
2Y^{}_{1} && -Y^{}_{3} && -Y^{}_{2} \\
-Y^{}_{3} && 2Y^{}_{2} && -Y^{}_{1} \\
-Y^{}_{2} && -Y^{}_{1} && 2Y^{}_{3}
\end{matrix}\right)\left.\rule{0cm}{1cm}\right]^{\ast}_{} \; .
\label{eq:mnu}
\end{eqnarray}
From Eq.~(\ref{eq:mnu}) we can find that each element in the mass matrix $M^{}_{\rm II}$ contains only one single modular form with a weight of 2, whereas every element in the matrix generated by $M^{}_{\rm D}M^{\rm T}_{\rm D}$ is a multiplication of two modular forms of weight 4, thus having a total modular weight of 8. However, $M^{-1}_{\rm R}$ contributes another weight of $-6$, therefore the  overall weight of $M^{}_{\rm I}$ is also 2, same as the weight of $M^{}_{\rm II}$. 

\section{Low-energy Phenomenology} \label{sec:lowphe}
Next we discuss the low-energy phenomenology of our model. As can be seen from the previous section, there are totally eight free model parameters, which are a complex modulus $\tau$ (or equivalently two real parameters ${\rm Re}\,\tau$ and ${\rm Im}\,\tau$), three real parameters $v^{}_{\rm d}\alpha^{}_{3}/\sqrt{2}$, $\alpha^{}_{1}/\alpha^{}_{3}$ and $\alpha^{}_{2}/\alpha^{}_{3}$ in the charged-lepton sector and two real parameters $g$ and $\phi^{}_{g}$ as well as an overall factor $v^{2}_{\rm u} g^{2}_{1}/(2\Lambda)$ in the neutrino sector. The number of free parameters is the same as that of the low-energy observables. As a result, our model should be predictive. Then we proceed to explore the phenomenological implications for lepton mass spectra, flavor mixing and CP violation. We carry out a numerical analysis of our model and demonstrate that the predictions are consistent with the experimental data only in the NO case at the $1\sigma$ level. The main strategy for numerical analysis is analogous to what we have done in Ref.~\cite{Wang:2019ovr}. Here we list it as follows.
\begin{itemize}
	\item First of all, the modulus parameter $\tau$ is randomly generated in the right-hand part of the fundamental domain ${\cal G}$, which is defined as
	\begin{eqnarray}
	{\cal G} = \left\{ \tau \in \mathbb{C}: \quad {\rm Im}\,\tau > 0, \; | {\rm Re}\,\tau| \leq 0.5, \; |\tau| \geq 1 \right\} \; .
	\label{eq:fundo}
	\end{eqnarray}
	This domain can be identified by using the basic properties of the modular forms as clearly explained in Ref.~\cite{Novichkov:2018ovf}. One can also notice that if the replacement $\tau \to -\tau^{\ast}_{}$ is made in Eq.~(\ref{eq:Y3q}), the modular forms $Y^{}_i(\tau)$ will change to their complex-conjugate counterparts, i.e., $Y^{}_i(-\tau^{\ast_{}}) = Y^{\ast}_{i}(\tau)$. If we further replace $\widetilde{g}$ with $\widetilde{g}^{\ast}$ in the neutrino sector, all the lepton mass matrices will become their complex-conjugate counterparts. Under such a transformation, the theoretical predictions for all the experimental observables keep unchanged except that the signs of all CP-violating phases should be reversed. Therefore in this paper we only consider the domain where $0 \leq | {\rm Re}\,\tau| \leq 0.5$, while the predictions for mixing parameters in the left-hand part of the fundamental domain where $-0.5 \leq | {\rm Re}\,\tau| \leq 0$ can be simply obtained by changing the overall signs of all the CP-violating phases.
	\begin{table}[t]
		\begin{center}
			\vspace{-0.25cm} \caption{The best-fit values, the
				1$\sigma$ and 3$\sigma$ intervals, together with the values of $\sigma^{}_{i}$ being the symmetrized $1\sigma$ uncertainties, for three neutrino mixing angles $\{\theta^{}_{12}, \theta^{}_{13}, \theta^{}_{23}\}$, two mass-squared differences $\{\Delta m^2_{21}, \Delta m^2_{31}~{\rm or}~\Delta m^2_{32}\}$ and the Dirac CP-violating phase $\delta$ from a global-fit analysis of current experimental data~\cite{Esteban:2018azc,nufit4.1}.} \vspace{0.5cm}
			\begin{tabular}{c|c|c|c|c}
				\hline
				\hline
				Parameter & Best fit & 1$\sigma$ range &  3$\sigma$ range & $\sigma^{}_{i}$ \\
				\hline
				\multicolumn{5}{c}{Normal neutrino mass ordering
					$(m^{}_1 < m^{}_2 < m^{}_3)$} \\ \hline
				$\sin^2_{}\theta^{}_{12}$
				& $0.310$ & 0.298 --- 0.323 &  0.275 --- 0.350 & 0.0125 \\
				$\sin^2_{}\theta^{}_{13}$
				& $0.02241$ & 0.02176 --- 0.02307 &  0.02046 --- 0.02440  & 0.000655 \\
				$\sin^2_{}\theta^{}_{23}$
				& $0.558$  & 0.525 --- 0.578 &  0.427 --- 0.609  & 0.0265 \\
				$\delta~[^\circ]$ &  $222$ & 194 --- 260 &  141 --- 370 & 33 \\
				$\Delta m^2_{21}~[10^{-5}~{\rm eV}^2]$ &  $7.39$ & 7.19 --- 7.60 & 6.79 --- 8.01 & 0.205 \\
				$\Delta m^2_{31}~ [10^{-3}~{\rm eV}^2]$ &  $+2.523$ & +2.493 --- +2.555 & +2.432 --- +2.618 & 0.031 \\\hline
				\multicolumn{5}{c}{Inverted neutrino mass ordering
					$(m^{}_3 < m^{}_1 < m^{}_2)$} \\ \hline
				$\sin^2_{}\theta^{}_{12}$
				& $0.310$ & 0.298 --- 0.323 &  0.275 --- 0.350 & 0.0125\\
				$\sin^2_{}\theta^{}_{13}$
				& $0.02261$ & 0.02197 --- 0.02328 &  0.02066 --- 0.02461 & 0.000655 \\
				$\sin^2_{}\theta^{}_{23}$
				& $0.563$  & 0.537 --- 0.582 &  0.430 --- 0.612 & 0.0225  \\
				$\delta~[^\circ]$ &  $285$ & 259 --- 309 &  205 --- 354 & 25 \\
				$\Delta m^2_{21}~[10^{-5}~{\rm eV}^2]$ &  $7.39$ & 7.19 --- 7.60 & 6.79 --- 8.01 & 0.205 \\
				$\Delta m^2_{32}~[10^{-3}~{\rm eV}^2]$ &  $-2.509$ & $-2.539$ --- $-2.477$  & $-2.603$ --- $-2.416$ & $0.031$ \\ \hline\hline
			\end{tabular}
			\label{table:gfit}
		\end{center}
	\end{table}
	
	In the charged-lepton sector, once we randomly choose the values of $\{ {\rm Re}\,\tau, {\rm Im}\,\tau\}$, the parameters $v^{}_{\rm d} \alpha^{}_{3}/\sqrt{2}$, $\alpha^{}_{1}/\alpha^{}_{3}$ and $\alpha^{}_{2}/\alpha^{}_{3}$ can be calculated from the following identities
	\begin{eqnarray}
	{\rm Tr} \left( M^{}_l M^{\dag}_l \right) &=& m^{2}_{e} + m^{2}_{\mu} + m^{2}_{\tau} \; ,  \label{eq:tr}\\
	{\rm Det}\left( M^{}_l M^{\dag}_l \right) &=& m^{2}_{e} m^{2}_{\mu} m^{2}_{\tau} \; , \label{eq:det}\\
	\dfrac{1}{2}\left[{\rm Tr} \left(M^{}_l M^{\dag}_l\right)\right]^2_{} - \dfrac{1}{2}{\rm Tr}\left[ (M^{}_l M^{\dag}_l)^2_{}\right] &= & m^{2}_{e}m^{2}_{\mu}+m^{2}_{\mu}m^{2}_{\tau}+m^{2}_{\tau}m^{2}_{e} \; , \label{eq:tr2}
	\end{eqnarray}
	where we take $m^{}_e=0.511~{\rm MeV}$, $m^{}_\mu=105.7~{\rm MeV}$ and $m^{}_\tau=1776.86~{\rm MeV}$  for the observed charged-lepton masses~\cite{Zyla:2020}. Notice that Eqs.~(\ref{eq:tr})-(\ref{eq:tr2}) have multiple solutions, corresponding to the different hierarchies of $\alpha^{}_{1}$, $\alpha^{}_{2}$ and $\alpha^{}_{3}$. Later we will see there exist two kinds of hierarchies ($\alpha^{}_{1} \ll \alpha^{}_{3} \ll \alpha^{}_{2}$ and $\alpha^{}_{1} \ll \alpha^{}_{2} \ll \alpha^{}_{3}$) which can lead to the realistic mixing pattern, with different predictions to the value of $\theta^{}_{23}$. So far all the parameters in $M^{}_l$ have been determined. It is then easy to diagonalize the charged-lepton mass matrix via $U^{\dag}_l M^{}_l M^{\dag}_l U^{}_l = {\rm Diag} \left\{ m^{2}_{e}, m^{2}_{\mu}, m^{2}_{\tau}\right\}$, from which the unitary matrix $U^{}_l$ can be obtained.

	\item Next the values of the other two parameters $g \in (0,10]$ and $\phi^{}_{g} \in [0^{\circ}_{}, 360^{\circ}_{})$ are randomly generated. Therefore, the effective neutrino mass matrix $M^{}_{\nu}$ is determined up to the overall scale parameter $v^2_{\rm u} g^2_1/(2\Lambda)$. We introduce a ratio $r$ defined as $r \equiv \Delta m^{2}_{21}/\Delta m^{2}_{31}$ in the NO case or $r \equiv \Delta m^{2}_{21}/|\Delta m^{2}_{32}|$ in the IO case which does not depend on this overall scale parameter, and this ratio can help us restrict the values of ${\rm Re}\,\tau$, ${\rm Im}\,\tau$, $g$ and $\phi^{}_g$. The overall parameter $v^2_{\rm u} g^2_1/(2\Lambda)$ can be determined right after we fix the value of the lightest neutrino mass. After diagonalizing $M^{}_\nu$ via $U^{\dag}_{\nu} M^{}_{\nu} U^{\ast}_{\nu} = {\rm Diag}\left\{m^{}_{1}, m^{}_{2}, m^{}_{3}\right\}$, we get the unitary matrix $U^{}_{\nu}$. Finally, the lepton flavor mixing matrix $U = U^{\dag}_l U^{}_{\nu}$ can be calculated by using $U^{}_l$ and $U^{}_\nu$. In the standard parametrization~\cite{Zyla:2020}, we have
	\begin{eqnarray}
	U = \left(
	\begin{matrix}
	c^{}_{12}c^{}_{13} && s^{}_{12}c^{}_{13} && s^{}_{13}e^{-{\rm i}\delta}_{} \\
	-s^{}_{12}c^{}_{23}-c^{}_{12}s^{}_{23}s^{}_{13}e^{{\rm i}\delta} && c^{}_{12}c^{}_{23}-s^{}_{12}s^{}_{23}s^{}_{13}e^{{\rm i}\delta}_{} && s^{}_{23}c^{}_{13} \\
	s^{}_{12}s^{}_{23}-c^{}_{12}c^{}_{23}s^{}_{13}e^{{\rm i}\delta} && -c^{}_{12}s^{}_{23}-s^{}_{12}c^{}_{23}s^{}_{13}e^{{\rm i}\delta} &&
	c^{}_{23}c^{}_{13}
	\end{matrix}\right)\left(
	\begin{matrix}
	e^{{\rm i}\rho} && ~  && ~\\
	~ && e^{{\rm i}\sigma} && ~ \\
	~ && ~ && 1
	\end{matrix}\right) \; ,
	\label{eq:UPMNS}
	\end{eqnarray}
	where $c^{}_{ij} \equiv \cos \theta^{}_{ij}$ and $s^{}_{ij} \equiv \sin \theta^{}_{ij}$ (for $ij =12,13,23$) have been defined and $\delta$, $\rho$ and $\sigma$ are the Dirac and two Majorana CP-violating phases, respectively. 
\end{itemize}
\begin{figure}[t!]
	\centering		\includegraphics[width=0.98\textwidth]{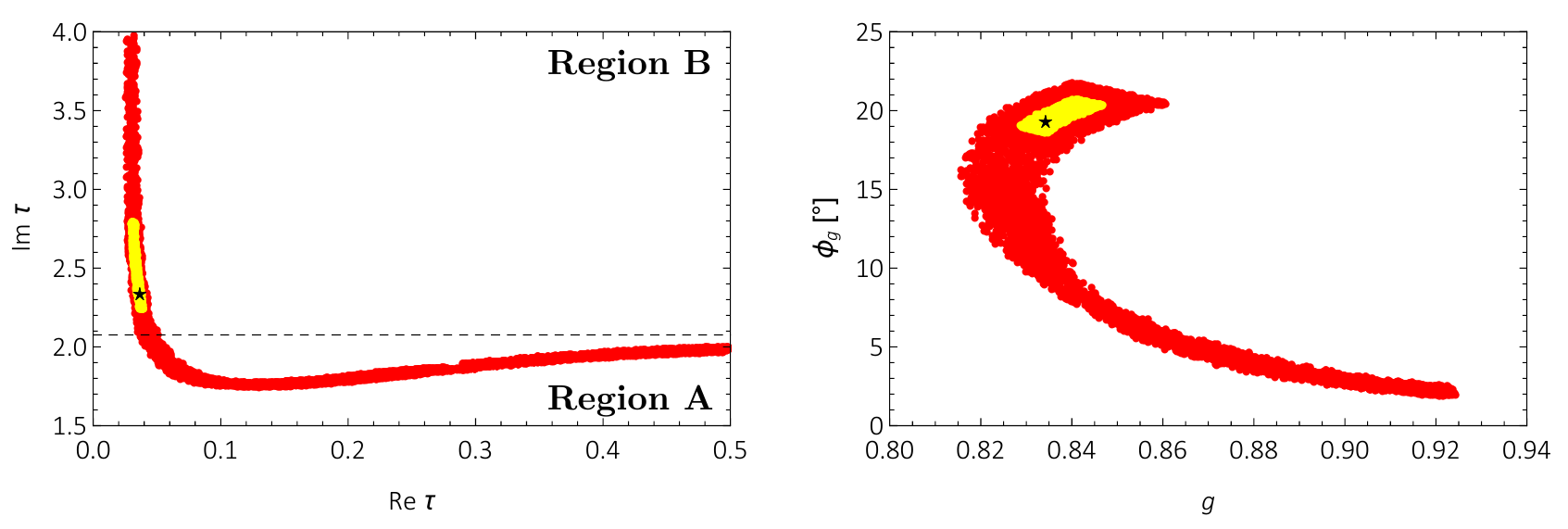}
	\vspace{0cm}
	\caption{Allowed ranges of the model parameters $\{{\rm Re}\,\tau, {\rm Im}\,\tau\}$ and $\{g, \phi^{}_g\}$ in the NO case, where the $1\sigma$ (yellow dots) and $3\sigma$ (red dots) ranges of neutrino mixing parameters and mass-squared differences from the global-fit analysis of neutrino oscillation data have been input~\cite{Esteban:2018azc,nufit4.1}. The best-fit values are indicated by the black stars. In the left panel, the horizontal dashed line separates the parameter space of $\{{\rm Re}\,\tau, {\rm Im}\,\tau\}$ into two regions: {\bf Region A} and {\bf Region B}.}
	\label{fig:freepara} 
\end{figure}
\begin{figure}[t!]
	\centering		\includegraphics[width=0.98\textwidth]{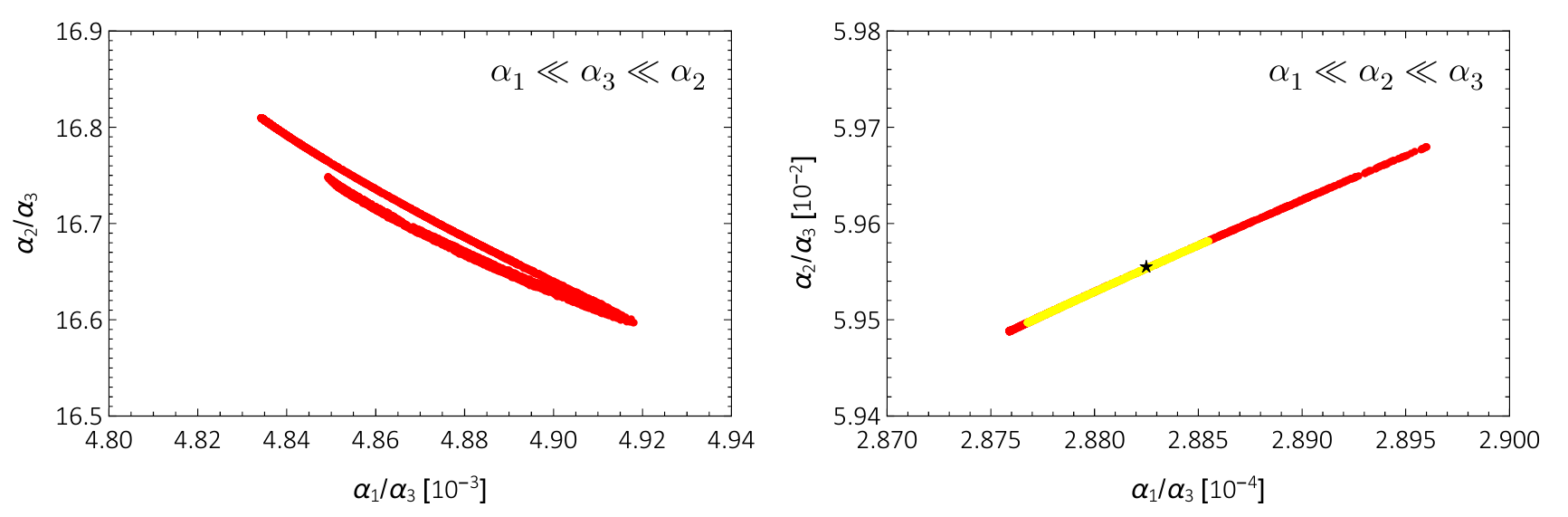}
\vspace{0cm}
	\caption{Allowed ranges of two ratios $\{\alpha^{}_{1}/\alpha^{}_{3},\alpha^{}_{2}/\alpha^{}_{3}\}$ in the charged-lepton sector for the NO case, where the $1\sigma$ (yellow dots) and $3\sigma$ (red dots) ranges  of neutrino mixing parameters and mass-squared differences from the global-fit analysis of neutrino oscillation data have been input~\cite{Esteban:2018azc,nufit4.1}. The best-fit value is indicated by the black star. The left panel corresponds to the hierarchy $\alpha^{}_{1} \ll \alpha^{}_{3} \ll \alpha^{}_{2}$ where only the $3\sigma$ range is allowed and the right panel is related to $\alpha^{}_{1} \ll \alpha^{}_{2} \ll \alpha^{}_{3}$. }
	\label{fig:abc} 
\end{figure}

To find out the allowed parameter space, we implement the global-fit results from NuFIT 4.1~\cite{Esteban:2018azc,nufit4.1} without including the atmospheric neutrino data from Super-Kamiokande. The best-fit values of three neutrino mixing angles $\{\theta^{}_{12}, \theta^{}_{13}, \theta^{}_{23}\}$, two neutrino mass-squared differences $\{\Delta m^2_{21}, \Delta m^2_{31}\}$ (or $\{\Delta m^2_{21}, \Delta m^2_{32}\}$), the Dirac CP-violating phase $\delta$, together with their $1\sigma$ and $3\sigma$ ranges in the NO (or IO) case are summarized in Table~\ref{table:gfit}. The $1\sigma$ ($3\sigma$) allowed parameter space can be determined by identifying whether the values of observables predicted by our model are located in their individual $1\sigma$ ($3\sigma$) ranges from the global analysis.

As we have mentioned before, the predictions of our model are consistent with the global-fit results in the NO case at the $1\sigma$ level. While in the IO case, we find that if all the other low-energy observables are within their individual $3\sigma$ ranges, the value of $\sin^{2}_{}\theta^{}_{23}$ will be either larger than 0.75 or very close to 0, both of which have already exceeded the $3\sigma$ range of $\sin^{2}_{}\theta^{}_{23}$. Therefore, the IO case is not compatible with current experimental data at the $3\sigma$ level. 

The allowed parameter space of $\{{\rm Re}\,\tau, {\rm Im}\,\tau\}$,  $\{g, \phi^{}_g\}$ and $\{\alpha^{}_{1}/\alpha^{}_{3},\alpha^{}_{2}/\alpha^{}_{3}\}$ has been shown in Figs.~\ref{fig:freepara}-\ref{fig:abc}, where the $1\sigma$ ($3\sigma$) range is denoted by the yellow (red) dots. As one can see from the left panel of Fig.~\ref{fig:freepara}, almost all the range $[0,0.5]$ of ${\rm Re}\,\tau$ is allowed at the $3\sigma$ level while the value of ${\rm Im}\,\tau$ is restricted to be larger than 1.75. Note that here we artificially cut off the parameter space of ${\rm Im}\,\tau$ at ${\rm Im}\,\tau = 4$, since in the range where ${\rm Im}\,\tau>4$, we find that the predicted values of mixing angles and CP-violating phases tend to be stable and the sum of three light neutrino masses $\sum m^{}_{\nu} = m^{}_{1}+m^{}_{2}+m^{}_{3}>2~{\rm eV}$, which has already been far away from the favored region of the latest Planck observations~\cite{Aghanim:2018eyx}, thus being out of our interest. Actually we can separate the parameter space of $\{{\rm Re}\,\tau, {\rm Im}\,\tau\}$ into two regions depending on the values of ${\rm Im}\,\tau$, to be specific, {\bf Region A} with $1.75<{\rm Im}\,\tau<2.07$ and {\bf Region B} with $2.07<{\rm Im}\,\tau<4$. An important feature to distinguish these two regions is that only the hierarchy $\alpha^{}_{1} \ll \alpha^{}_{3} \ll \alpha^{}_{2}$ is permitted in {\bf Region A} while both the hierarchies $\alpha^{}_{1} \ll \alpha^{}_{3} \ll \alpha^{}_{2}$ and $\alpha^{}_{1} \ll \alpha^{}_{2} \ll \alpha^{}_{3}$ are allowed in {\bf Region B}. The reason for this fact will be discussed in detail later. On the other hand, in {\bf Region A}  the value of ${\rm Im}\,\tau$ can only change in a narrow region. However ${\rm Re}\,\tau$ can vary in a wide range, from 0.04 to 0.5. On the contrary, in {\bf Region B} the value of ${\rm Re}\,\tau$ is about 0.03 while ${\rm Im}\,\tau$ can reach very large values. The constraints on $g$ and $\phi^{}_g$ within the $3\sigma$ level are $0.82<g<0.92$ and $1.92^{\circ}_{}<\phi^{}_{g}<21.8^{\circ}_{}$ respectively, as can be seen from the right panel of Fig.~\ref{fig:freepara}. The value of $\widetilde{g}$ measures the individual contributions to the neutrino masses from the type-I and type-II seesaw mechanisms, and $g \sim 1$ means that their individual contributions are comparable to each other. 

\begin{figure}[t!]
	\centering		
	\includegraphics[width=0.49\textwidth]{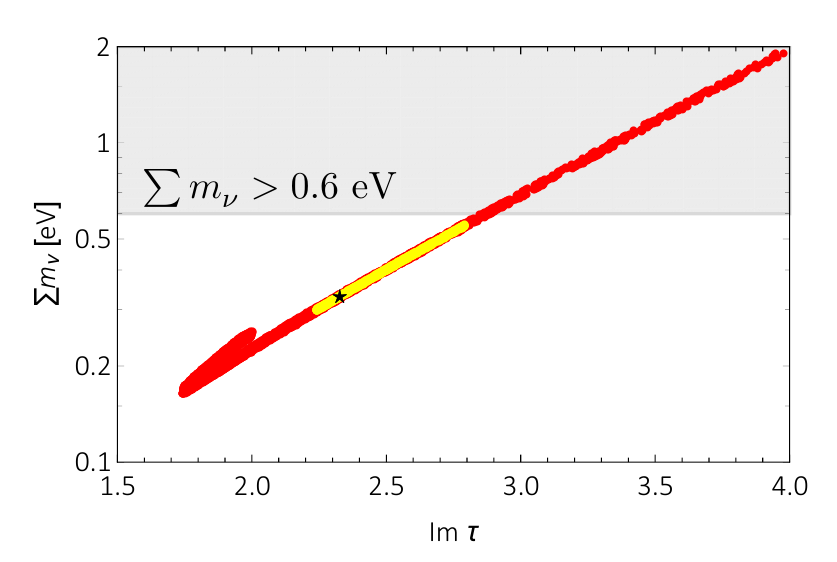}
	\vspace{0cm}
	\caption{The relation between the sum of light neutrino masses $\sum m^{}_{\nu}=m^{}_1+m^{}_2+m^{}_3$  and ${\rm Im}\,\tau$ for the NO case, where the $1\sigma$ (yellow dots) and $3\sigma$ (red dots) ranges of neutrino mixing parameters and mass-squared differences from the global-fit analysis of neutrino oscillation data have been input~\cite{Esteban:2018azc,nufit4.1}. The best-fit value is indicated by the black star. The gray shaded region denotes the range $\sum m^{}_{\nu}>0.6~{\rm eV}$ which is excluded by the latest Planck observations~\cite{Aghanim:2018eyx}.}
	\label{fig:m123} 
\end{figure}

To determine the model parameters from neutrino oscillation data and describe how well the model is consistent with observations, we construct the $\chi^2$-function by regarding the best-fit values $q^{\rm bf}_j$ of the oscillation parameters $q^{}_j \in \{\sin^2\theta^{}_{12}, \sin^2\theta^{}_{13}, \sin^2\theta^{}_{23}, \Delta m^2_{21}, \Delta m^2_{31}\}$ from the global analysis in Refs.~\cite{Esteban:2018azc,nufit4.1} as experimental measurements, namely,
\begin{eqnarray}
\chi^2_{}(p^{}_{i}) = \sum^{}_{j}\left(\frac{q^{}_{j}(p^{}_{i})-q^{\rm bf}_{j}}{\sigma^{}_{j}}\right)^2_{} \; ,
\label{eq:chi}
\end{eqnarray}
where $p^{}_{i} \in \{{\rm Re}\,\tau, {\rm Im}\,\tau, g, \phi^{}_g, v^2_{\rm u}g^2_1/(2\Lambda)\}$ stand for the free model parameters and $q^{}_j(p^{}_i)$ denote the model predictions for observables with $\sigma^{}_j$ being the symmetrized $1\sigma$ uncertainties from the global-fit analysis, which has already been given in Table~\ref{table:gfit}. Since the current constraint on $\delta$ from the global-fit results is rather weak, we will not include the information of $\delta$ in the $\chi^2_{}$-function. Note that in this work, the $\chi^2$-function is only used to determine the best-fit values of free model parameters. Based on the $\chi^2$-fit analysis, we find that the minimum $\chi^{2}_{\rm min} = 0.0862$ is obtained in the NO case with the following best-fit values of the model parameters
\begin{eqnarray}
{\rm Re}\,\tau = 0.0366\;, \quad {\rm Im}\,\tau = 2.33 \;, \quad g = 0.834 \;, \quad \phi^{}_{g} = 19.2^{\circ}_{} \;, \quad v^2_{\rm u}g^2_1/(2\Lambda)=0.1344~{\rm eV} \; ,
\label{eq:bfBN1}
\end{eqnarray}
which together with the charged-lepton masses $m^{}_\alpha$ (for $\alpha = e, \mu, \tau$) lead to $v^{}_{\rm d} \alpha^{}_3 /\sqrt{2} = 1.775~{\rm GeV}$, $\alpha^{}_1/\alpha^{}_3 = 2.88\times 10^{-4}$ and $\alpha^{}_2/\alpha^{}_3 = 5.96\times 10^{-2}$. Once the best-fit values of model parameters are known, one can easily obtain the best-fit values of the observables $q^{}_j$ and the predictions for the CP-violating phases $\delta$, $\rho$ and $\sigma$, as well as the effective neutrino mass $m^{}_\beta$ for beta decays and $m^{}_{\beta\beta}$ for neutrinoless double-beta decays, which can be found in Table~\ref{table:value}.

The relation between $\sum m^{}_{\nu}$ and ${\rm Im}~\tau$ is presented in Fig.~\ref{fig:m123}, from which we can find that the value of $\sum m^{}_{\nu}$ is tightly related to ${\rm Im}\,\tau$. To be specific, as the value of ${\rm Im}\,\tau$ increases, $\sum m^{}_{\nu}$ will also increase. This can be understood in the following way. Since ${\rm Im}\,\tau$ in our model is larger than 1.75, $|q|=e^{-2\pi{\rm Im}\,\tau}<4\times 10^{-3}_{}$ in the Fourier expansions of modular forms $Y^{}_{i}(\tau)$ (for $i=1,2,3$) in Eq.~(\ref{eq:Y3q}) is a small parameter. Therefore we can retain only the leading order terms in the expansions of $Y^{}_{i}(\tau)$ and Eq.~(\ref{eq:Y3q}) will change to
\begin{eqnarray}
Y^{}_{1} \approx 1 \; , \quad Y^{}_{2} \approx -6q^{1/3}_{} \equiv t\; , \quad Y^{}_{3} \approx -18q^{2/3}_{} \equiv -\frac{1}{2}t^{2}_{} \; ,
\label{eq:approY}
\end{eqnarray}
where $t \equiv -6q^{1/3}_{} = -6 e^{2\pi{\rm i}\tau/3}$ has been defined. Then the neutrino mass matrix $M^{}_{\nu}$ can be expressed in an approximate form up to ${\cal O}(t^3_{})$
\begin{figure}[t!]
	\centering		
	\includegraphics[width=0.98\textwidth]{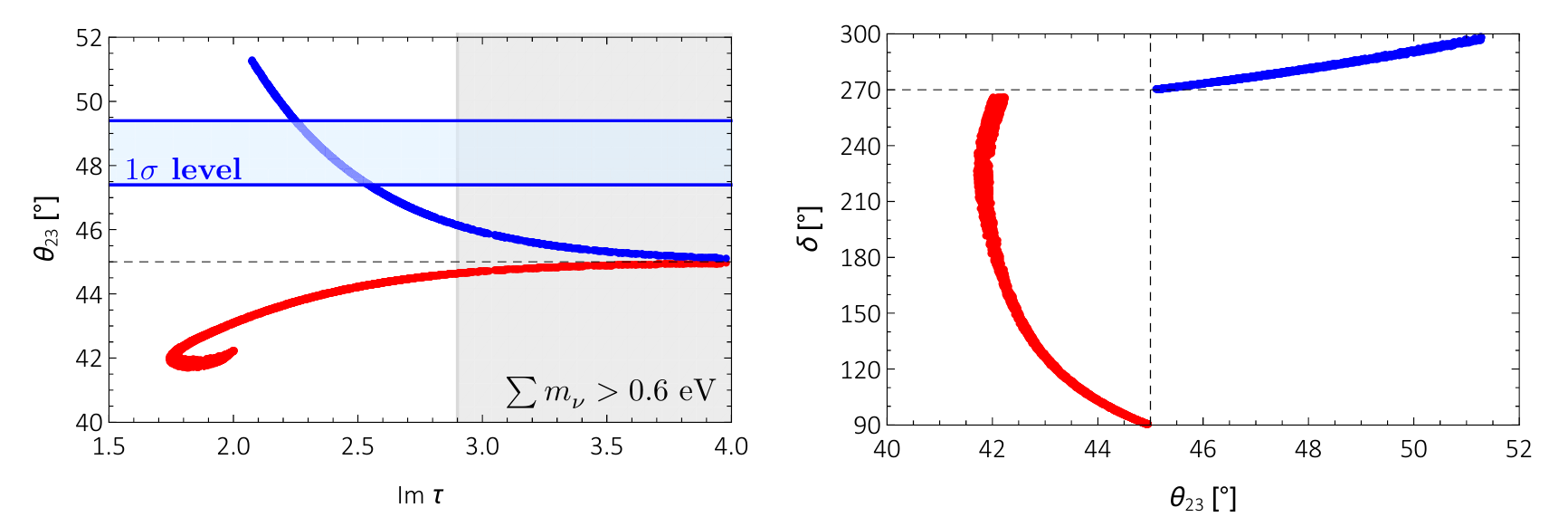}
	\vspace{0cm}
	\caption{In the left panel, the correlation between $\theta^{}_{23}$ and ${\rm Im}\,\tau$ within the $3\sigma$ level is shown for the NO case while the correlation between $\delta$ and $\theta^{}_{23}$ is presented in the right panel. The red and blue curves correspond to the hierarchies $\alpha^{}_1 \ll \alpha^{}_3 \ll \alpha^{}_2$ and $\alpha^{}_1 \ll \alpha^{}_2 \ll \alpha^{}_3$, respectively. The gray shaded region in the left panel represents the range $\sum m^{}_{\nu}>0.6~{\rm eV}$ which is excluded by the latest Planck observations~\cite{Aghanim:2018eyx} while the blue shaded region denotes the range of $\theta^{}_{23}$ within the $1\sigma$ level. The horizontal (vertical) dashed line in the left (right) panel refers to $\theta^{}_{23}=45^{\circ}_{}$ while the horizontal dashed line in the right panel denotes $\delta=270^{\circ}_{}$.}
	\label{fig:theta231} 
\end{figure}
\begin{eqnarray}
M^{}_{\nu} \approx -\frac{v^{2}_{\rm u} g^{2}_{1}}{2\Lambda}\left(\begin{matrix}
1-2\widetilde{g}-\dfrac{3}{2}t^3_{} && \dfrac{1}{2}(3-\widetilde{g})t^2_{} &&  -(1-\widetilde{g})t \\
\dfrac{1}{2}(3-\widetilde{g})t^2_{} && -2\widetilde{g}t && \widetilde{g}-\dfrac{3}{2}t^3_{} \\
-(1-\widetilde{g})t  && \widetilde{g}-\dfrac{3}{2}t^3_{} && (1+\widetilde{g})t^2_{}
\end{matrix}\right)^{\ast}_{} \; .
\label{eq:mnuappro}
\end{eqnarray}
All the elements except $(M^{}_{\nu})^{}_{11}$, $(M^{}_{\nu})^{}_{23}$ and $(M^{}_{\nu})^{}_{32}$ in the right-hand side of Eq.~(\ref{eq:mnuappro}) are suppressed by the higher order terms of $t$. As the value of ${\rm Im}\,\tau$ becomes larger, $(M^{}_{\nu})^{}_{11}$, $(M^{}_{\nu})^{}_{23}$ and $(M^{}_{\nu})^{}_{32}$ will dominant the eigenvalues of $M^{}_{\nu}$. Given the parameter space of $\widetilde{g}$, we can find that the modulus of $(M^{}_{\nu})^{}_{11}$ is close to that of $(M^{}_{\nu})^{}_{23}$ (Note that $(M^{}_{\nu})^{}_{23}=(M^{}_{\nu})^{}_{32}$ exactly holds due to the nature of the Majorana mass matrix) especially when ${\rm Im}\,\tau$ is large enough, implying a quasi-degeneracy among $m^{}_{1}$, $m^{}_{2}$ and $m^{}_{3}$. The high degeneracy of three neutrino masses requires a large $\sum m^{}_{\nu}$, which is why the value of $\sum m^{}_{\nu}$ increases with the rise of ${\rm Im}\,\tau$.

Fig.~\ref{fig:m123} also indicates that the minimal value of $\sum m^{}_{\nu}$ predicted in our model is $0.16~{\rm eV}$, which has already exceeded the upper bound on the sum of neutrino masses $\sum m^{}_{\nu} < 0.12~{\rm eV}$ from the Planck observations~\cite{Aghanim:2018eyx}. Notice that the limit $\sum m^{}_{\nu} < 0.12~{\rm eV}$ has also been obtained earlier in Refs.~\cite{Giusarma:2016phn,Vagnozzi:2017ovm}. However this upper bound is cosmological model dependent and obtained by combining other experimental data such as the baryon acoustic oscillation (BAO), the gravitational lensing of galaxies and the high multipole TT, TE and EE polarization spectra. If only the BAO data and the cosmic microwave background (CMB) lensing reconstruction power spectrum are taken into consideration, the restriction to $\sum m^{}_{\nu}$ can be liberalized to $\sum m^{}_{\nu}<0.6~{\rm eV}$~\cite{Aghanim:2018eyx}. Therefore, our model can still be compatible with the Planck observations in the region where $0.12~{\rm eV}<\sum m^{}_{\nu}<0.6~{\rm eV}$. In addition, since our model predicts relatively large values of three neutrino masses, it can be easily tested in future neutrino experiments.

A salient feature of our model is that the predicted value of $\theta^{}_{23}$ shows a strong dependence on the free model parameters especially ${\rm Im}\,\tau$, as can be seen from the left panel of Fig.~\ref{fig:theta231}, where the red and blue curves correspond to the hierarchy $\alpha^{}_1 \ll \alpha^{}_3 \ll \alpha^{}_2$ and $\alpha^{}_1 \ll \alpha^{}_2 \ll \alpha^{}_3$, respectively. Some useful remarks are as follows.
\begin{itemize}
	\item There are two branches in the allowed range of $\{{\rm Im}\,\tau,\theta^{}_{23}\}$, depending on which hierarchy of $\alpha^{}_1$, $\alpha^{}_2$ and $\alpha^{}_3$ is taken into consideration. In the hierarchy $\alpha^{}_1 \ll \alpha^{}_3 \ll \alpha^{}_2$, $\theta^{}_{23}$ is located in the first octant where $\theta^{}_{23}<45^{\circ}_{}$ while in the hierarchy $\alpha^{}_1 \ll \alpha^{}_2 \ll \alpha^{}_3$, $\theta^{}_{23}$ is in the second octant, which 
	is preferred by the latest global-fit analysis within $1\sigma$ level. If further long baseline experiments such as DUNE~\cite{Acciarri:2015uup} and Hyper-Kamiokande~\cite{Abe:2018uyc} can give more precise measurements on the octant of $\theta^{}_{23}$, it will be promising to determine the hierarchy of $\alpha^{}_1$, $\alpha^{}_2$ and $\alpha^{}_3$ in our model unambiguously.
	\item In order to illustrate how the value of $\theta^{}_{23}$ is connected with the hierarchies of $\alpha^{}_1$, $\alpha^{}_2$ and $\alpha^{}_3$, we express $M^{}_{l}$ in an approximate form by substituting Eq.~(\ref{eq:approY}) into Eq.~(\ref{eq:Me})
	\begin{eqnarray}
	M^{}_{l} \approx \frac{v^{}_{\rm d}}{\sqrt{2}}\left(
	\begin{matrix}
	1 && t && -\dfrac{1}{2}t^2_{} \\
	-\dfrac{1}{2}t^2_{} && 1 && t \\
	t && -\dfrac{1}{2}t^2_{} && 1 
	\end{matrix}
	\right)^{\ast}_{}
	\left(
	\begin{matrix}
	\alpha^{}_{1} && 0 && 0 \\
	0 && \alpha^{}_{2} && 0  \\
	0 && 0 && \alpha^{}_{3} 
	\end{matrix}
	\right) \; .
	\label{eq:Meappro}
	\end{eqnarray}
	Since $t$ can be regarded as a small parameter and there is a strong hierarchy of $\alpha^{}_1$, $\alpha^{}_2$ and $\alpha^{}_3$ (namely $\alpha^{}_1 \ll \alpha^{}_3 \ll \alpha^{}_2$ or $\alpha^{}_1 \ll \alpha^{}_2 \ll \alpha^{}_3$), it is possible to obtain the approximate analytical form of the unitary matrix $U^{}_{l}$. To be specific, in the hierarchy $\alpha^{}_1 \ll \alpha^{}_3 \ll \alpha^{}_2$, the unitary matrix $U^{(1)}_{l}$can be written as
	\begin{eqnarray}
	U^{(1)}_{l} \approx \left(
	\begin{matrix}
	-1+\dfrac{1}{2}|t|^2_{} && -\dfrac{3}{2}(t^{\ast}_{})^2_{} && t^{\ast}_{} \\
	t && \dfrac{1}{2}t^2_{} && 1-\dfrac{1}{2}|t|^2_{} \\
	-\dfrac{3}{2}t^2_{} && 1 && -\dfrac{1}{2}(t^{\ast}_{})^2_{}
	\end{matrix}\right) \; ,
	\label{eq:Ulappro1}
	\end{eqnarray}
	where we retain only the terms up to ${\cal O}(t^2_{})$. Then the lepton mixing matrix $U^{(1)}_{}$ turns out to be $U^{(1)}_{}=(U^{(1)}_{l})^{\dag}_{}U^{}_{\nu}$, and $\sin^2_{}\theta^{(1)}_{23}$ can be obtained from $\sin^2_{}\theta^{(1)}_{23}=|U^{(1)}_{\mu 3}|^2_{}/(1-|U^{(1)}_{e3}|^2_{})$. While in the hierarchy $\alpha^{}_1 \ll \alpha^{}_2 \ll \alpha^{}_3$, the unitary matrix $U^{(2)}_{l}$ is
	\begin{eqnarray}
	U^{(2)}_{l} \approx \left(
	\begin{matrix}
	-1+\dfrac{1}{2}|t|^2_{} && t^{\ast}_{} && -\dfrac{1}{2}(t^{\ast}_{})^2_{} \\
	t && 1-|t|^2_{} && t^{\ast}_{} \\
	-\dfrac{3}{2}t^2_{} && -t && 1-\dfrac{1}{2}|t|^2_{}
	\end{matrix}\right) \; .
	\label{eq:Ulappro2}
	\end{eqnarray}
	The lepton mixing matrix $U^{(2)}_{}$ and $\sin^2_{}\theta^{(2)}_{23}$ then can be expressed as $U^{(2)}_{}=(U^{(2)}_{l})^{\dag}_{}U^{}_{\nu}$ and $\sin^2_{}\theta^{(2)}_{23}=|U^{(2)}_{\mu 3}|^2_{}/(1-|U^{(2)}_{e3}|^2_{})$, respectively. We can use another unitary matrix $P$ to connect $U^{(1)}_{l}$ and $U^{(2)}_{l}$ via $U^{(2)}_{l} = U^{(1)}_{l} P$. Now that we have already obtained the approximate expressions of $U^{(1)}_{l}$ and $U^{(2)}_{l}$, it is easy to write down the explicit form of $P$
	\begin{eqnarray}
	P= \left(
	\begin{matrix}
	1 && 0 && 0 \\
	0 && -t+\dfrac{1}{2}(t^{\ast}_{})^2_{} && 1-\dfrac{1}{2}|t|^2_{} \\
	0 && 1-\dfrac{1}{2}|t|^2_{} && t^{\ast}_{}
	\end{matrix}\right) \; .
	\label{eq:P}
	\end{eqnarray}
    Then we can express $U^{(2)}_{e2}$, $U^{(2)}_{e3}$ and $U^{(2)}_{\mu 3}$ by using the elements of $U^{(1)}_{}$ as
	\begin{eqnarray}
	U^{(2)}_{e2}&=&U^{(1)}_{e2} \; ,  \nonumber \\
	U^{(2)}_{e3}&=&U^{(1)}_{e3} \; ,  \nonumber \\
	U^{(2)}_{\mu 3}&=& \left(-t^{\ast}_{}+\dfrac{1}{2}t^2_{}\right)U^{(1)}_{\mu 3}+ \left(1-\dfrac{1}{2}|t|^2_{} \right)U^{(1)}_{\tau 3} \; .
	\label{eq:u2ele}
	\end{eqnarray}
	So finally we arrive at the expression of $\sin^2_{}\theta^{(2)}_{23}$
	\begin{eqnarray}
	\sin^{2}_{}\theta^{(2)}_{23} = |t|^2_{}\sin^{2}_{}\theta^{(1)}_{23}+\left(1-\dfrac{1}{2}|t|^2_{}\right)\cos^2_{}\theta^{(1)}_{23}-\dfrac{2{\rm Re}\left[(t^{\ast}_{}-t^2_{}/2) U^{(1)}_{\mu 3}(U^{(1)}_{\tau 3})^{\ast}_{}\right]}{1-|U^{(1)}_{e 3}|^2_{}} \; .
	\label{eq:s232}
	\end{eqnarray}
	If ${\rm Im}\,\tau$ is sufficiently large, all the higher order terms of $t$ can be neglected, and we will arrive at $\sin^{2}_{}\theta^{(1)}_{23} \approx \cos^{2}_{}\theta^{(2)}_{23}$, i.e., $\theta^{(1)}_{23}+\theta^{(2)}_{23} \approx 90^{\circ}_{}$. However when ${\rm Im}\,\tau$ is not large enough, we should take the modification from higher order terms of $t$ into consideration, especially the term of ${\cal O}(t)$. The numerical calculation shows that ${\rm Re}\,[t^{\ast}_{} U^{(1)}_{\mu 3}(U^{(1)}_{\tau 3})^{\ast}_{}]<0$, indicating $\sin^{2}_{}\theta^{(2)}_{23}>\cos^2_{}\theta^{(1)}_{23}$, as can be seen from the left panel of Fig.~\ref{fig:theta231}. Let us consider a critical case where ${\rm Re}\,\tau=0.0398$, ${\rm Im}\,\tau=2.07$, $g=0.825$ and $\phi^{}_g=17.5^\circ_{}$. In the hierarchy $\alpha^{}_1 \ll \alpha^{}_3 \ll \alpha^{}_2$ we have
	\begin{eqnarray}
	\theta^{(1)}_{23} \simeq 43.3^{\circ}_{} \; , \quad \dfrac{2{\rm Re}\left[t^{\ast}_{} U^{(1)}_{\mu 3}(U^{(1)}_{\tau 3})^{\ast}_{}\right]}{1-|U^{(1)}_{e 3}|^2_{}} \simeq -0.0775 \; .
	\label{eq:h1value}
	\end{eqnarray} 
    Then from Eq.~(\ref{eq:s232}) we can obtain the value of $\theta^{(2)}_{23} \simeq 51.3^{\circ}_{}$, which is exactly the upper bound of the $3\sigma$ range from the global-fit results of $\theta^{}_{23}$. Therefore, if ${\rm Im}\,\tau<2.07$, we could not find the proper value of $\theta^{(2)}_{23}$ located in the $3\sigma$ range any more. That is why only the hierarchy $\alpha^{}_1 \ll \alpha^{}_3 \ll \alpha^{}_2$ is allowed in {\bf Region A}.
	\item The asymptotic behavior of $\theta^{}_{23}$ and $\delta$ when ${\rm Im}\,\tau$ is extremely large deserves some more discussion. As can be seen from Fig.~\ref{fig:theta231}, $\theta^{}_{23} \approx 45^{\circ}_{}$ and $\delta \approx 90^{\circ}_{}$ or $270^{\circ}_{}$ hold excellently when ${\rm Im}\,\tau \sim 4$. The distinctive values of $\theta^{}_{23}$ and $\delta$ imply that the neutrino mass matrix $M^{}_{\nu}$ might possess an approximate $\mu-\tau$ reflection symmetry~\cite{Harrison:2002et}, which means 
	\begin{eqnarray}
	X^{\rm T}_{\mu\tau}M^{}_{\nu}X^{}_{\mu\tau} = M^{\ast}_{\nu} \; ,
	\label{eq:mutaut}
	\end{eqnarray}
	where
	\begin{eqnarray}
	X^{}_{\mu\tau}=\left(
	\begin{matrix}
	1 && 0 && 0 \\
	0 && 0 && 1 \\
	0 && 1 && 0
	\end{matrix}\right) \; ,
	\label{eq:Xmutau}
	\end{eqnarray}
	 or equivalently $|U^{}_{\mu i}| = |U^{}_{\tau i}|$ (for $i=1,2,3$). Actually as we have mentioned before, except the $(M^{}_{\nu})^{}_{11}$, $(M^{}_{\nu})^{}_{23}$ and $(M^{}_{\nu})^{}_{32}$, all the other elements in Eq.~(\ref{eq:mnuappro}) are vanishing in the limit where ${\rm Im}\,\tau$ tends to infinity. Under this limit it is easy to identify Eq.~(\ref{eq:mutaut}) is satisfied up to a phase transition and $M^{}_{\nu}$ can be regarded as having a trivial $\mu-\tau$ reflection symmetry. However, the non-zero value of $t$ is required to slightly break down this symmetry and generate non-trivial values of the mass-squared differences as well as other mixing angles to fit the experiment data. In order to illustrate this approximate $\mu-\tau$ reflection symmetry indeed exists, we give a specific example of the numerical result for $|U^{}_{\nu}|$ as
	 \begin{eqnarray}
	 |U^{}_{\nu}| = \left(
	 \begin{matrix}
	 0.816 && 0.599 && 0.146 \\
	 0.408 && 0.586 && 0.700 \\
	 0.409 && 0.587 && 0.699
	 \end{matrix}\right) \; ,
	 \label{eq:Ununum}
	 \end{eqnarray}
	 where the free model parameters are set to be
	 \begin{eqnarray}
	 {\rm Re}\,\tau = 0.0275 \; , \quad {\rm Im}\,\tau = 3.95 \; , \quad g=0.856 \; , \quad \phi^{}_g=20.8^{\circ}_{} \; .
	 \label{eq:freeparanum}
	 \end{eqnarray}
	 Hence we can find $|U^{}_{\mu i}| \approx |U^{}_{\tau i}|$, which also indicates that there is an approximate $\mu-\tau$ reflection symmetry in $M^{}_{\nu}$.
	 \item As can be seen in Eq.~(\ref{eq:u2ele}), the elements $U^{}_{e 2}$ and $U^{}_{e 3}$ in the lepton mixing matrix $U$ keep invariant under the conversion from one hierarchy to another. Therefore different from $\theta^{}_{23}$, the values of $\theta^{}_{12}$ and $\theta^{}_{13}$ do not depend on the hierarchies of $\alpha^{}_{1}$, $\alpha^{}_{2}$ and $\alpha^{}_{3}$.  
\end{itemize} 
\begin{figure}[t!]
	\centering		
	\includegraphics[width=0.98\textwidth]{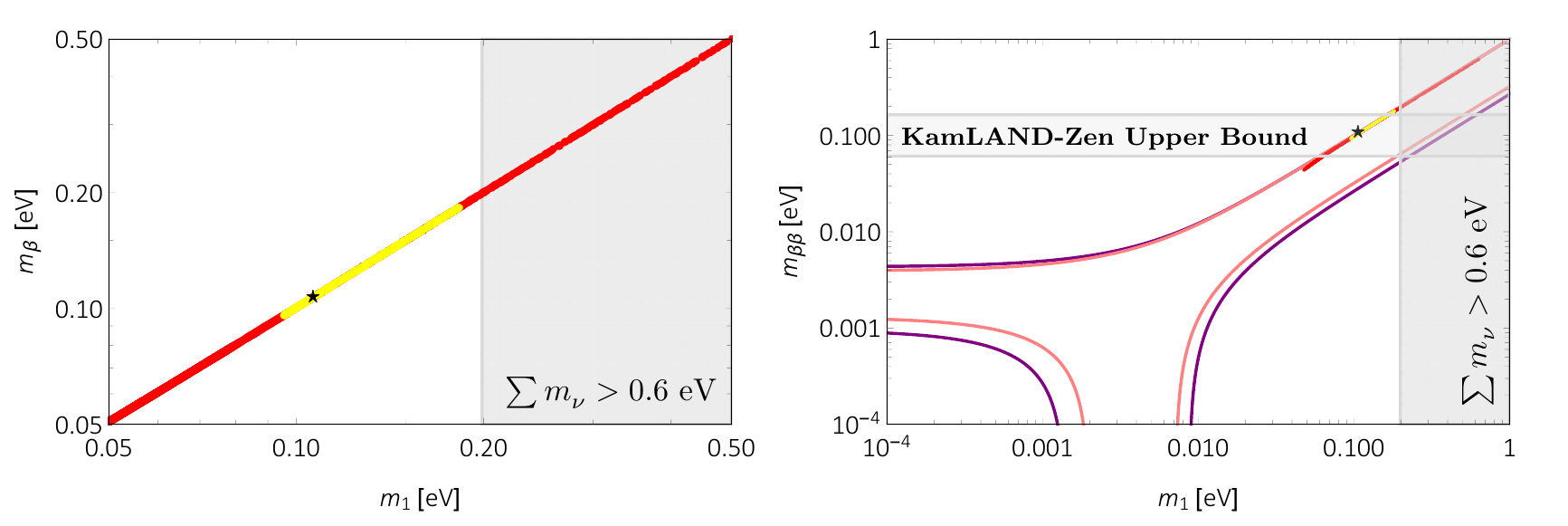}
	\vspace{0.cm}
	\caption{In the left panel, the correlation between $m^{}_{\beta}$ and $m^{}_1$ is shown for the NO case where the $1\sigma$ (yellow dots) and $3\sigma$ (red dots) ranges of neutrino mixing parameters and mass-squared differences from the global-fit analysis of neutrino oscillation data have been input~\cite{Esteban:2018azc,nufit4.1}. While the right panel is for the correlation between $m^{}_{\beta\beta}$ and $m^{}_1$. The gray shaded regions represent the range $\sum m^{}_{\nu}>0.6~{\rm eV}$ which is excluded by the latest Planck observations~\cite{Aghanim:2018eyx} while the light gray shaded region in the right panel denotes the upper bound on $m^{}_{\beta\beta}$ from the KamLAND-Zen experiment~\cite{KamLAND-Zen:2016pfg}. The pink (purple) boundary in the right panel is obtained by using the $1\sigma$ ($3\sigma$) ranges of $\{\theta^{}_{12},\theta^{}_{13}\}$ and $\{\Delta m^{2}_{21},\Delta m^{2}_{31}\}$ from the global-fit analysis.}
	\label{fig:effmass} 
\end{figure}

On the other hand, once the neutrino mass spectrum and the mixing parameters are known, we can predict the effective neutrino mass for beta decays,
\begin{eqnarray}
m^{}_\beta \equiv \sqrt{m^2_1 |U^{}_{e1}|^2 + m^2_2 |U^{}_{e2}|^2 + m^2_3 |U^{}_{e3}|^2} \; .
\label{eq:mbeta}
\end{eqnarray}
In the case where three neutrino masses are quasi-degenerate, $m^{}_{\beta}$ is approximately proportional to $m^{}_1$, as can be seen from the left panel of Fig.~\ref{fig:effmass}. The latest result from the KATRIN experiment, where the electron energy spectrum from tritium beta decays is precisely measured, indicates $m^{}_\beta < 1.1~{\rm eV}$ at the $90\%$ confidence level~\cite{Aker:2019uuj, Aker:2019qfn}. With more data accumulated in KATRIN, the upper bound will be improved to $m^{}_\beta < 0.2~{\rm eV}$. Then it will for sure provide us with some clues to test whether the corresponding parameter space in our model is still consistent with the experiment data or not.

Furthermore, three light neutrinos are Majorana particles in the seesaw model, indicating that the neutrinoless double-beta decays of some even-even heavy nuclei could take place. The effective neutrino mass relevant for neutrinoless double-beta decays is defined as
\begin{eqnarray}
m^{}_{\beta\beta} \equiv |m^{}_1 U^2_{e1} + m^{}_2 U^2_{e2} + m^{}_3 U^2_{e3}| \; .
\label{eq:mbb}
\end{eqnarray}
The right panel of Fig.~\ref{fig:effmass} shows that the predicted values of $m^{}_{\beta\beta}$ in our model have already reached the upper bound from the KamLAND-Zen experiment~\cite{KamLAND-Zen:2016pfg}, $m^{\rm upper}_{\beta\beta}=0.061$ -- $0.165~{\rm eV}$, which is currently the best experimental constraint on $m^{}_{\beta\beta}$. Hence our model is quite testable and can be easily ruled out in the next-generation neutrinoless double-beta decay experiments~\cite{Dolinski:2019nrj}.

\begin{table}[t]
	\begin{center}
		\vspace{-0.25cm} \caption{The best-fit values with the minimum $\chi^2_{\rm min}=0.0862$, together with the 1$\sigma$ and 3$\sigma$ ranges of all the free model parameters and observables for the NO case in our model.} \vspace{0.5cm}
		\begin{tabular}{c|c|c|c|c}
			\hline
			\hline
			  \multicolumn{2}{c|}{} & Best fit & $1\sigma$ range & $3\sigma$ range \\
			  \hline
			\multirow{8}*{\rotatebox[]{90}{Free model parameters}} 
			  & ${\rm Re}\,\tau$ & 0.0366 & 0.0304 -- 0.0394 & 0.0260 -- 0.5\\
			~ & ${\rm Im}\,\tau$ & 2.33 & 2.24 -- 2.80 & 1.75 -- 4 \\
			~ & $g$              & 0.834 & 0.829 -- 0.847 & 0.815 -- 0.924 \\ 
			~ & $\phi^{}_g$ $[{}^{\circ}_{}]$ & 19.2 & 18.6 -- 20.7 & 1.92 -- 21.8\\
			~ & $v^{}_{\rm d}\alpha^{}_{3}/\sqrt{2}$ $[{\rm GeV}]$ &  1.775 & 1.774 -- 1.777 & 0.1056 -- 0.1058 $\oplus$ 1.771 -- 1.777\\
			~ & $\alpha^{}_1/\alpha^{}_3$ $[10^{-4}_{}]$ & 2.88 & 2.88 -- 2.89 & 48.3 -- 49.2 $\oplus$ 2.88 -- 2.90 \\
			~ & $\alpha^{}_2/\alpha^{}_3$ & 0.0596 & 0.0595 -- 0.0596 & 16.6 -- 16.8 $\oplus$ 0.0595 -- 0.0597\\ 
			~ & $v^2_{\rm u}g^2_1/(2\Lambda)$ $[{\rm eV}]$ & 0.1344 & 0.1224 -- 0.2217 & 0.06874 -- 0.7776\\
			\hline
			\multirow{11}*{\rotatebox[]{90}{Observables}} 
			  & $m^{}_1$ $[{\rm eV}]$& 0.1066 & 0.09561 -- 0.1832 & 0.04778 -- 0.6592\\
			~ & $m^{}_2$ $[{\rm eV}]$& 0.1070 & 0.09600 -- 0.1834 & 0.04858 -- 0.6592\\
			~ & $m^{}_3$ $[{\rm eV}]$& 0.1178 & 0.1078 -- 0.1900 & 0.06875 -- 0.6611\\
			~ & $\theta^{}_{12}$ $[{}^{\circ}_{}]$ & 33.8 & 33.1 -- 34.6 & 31.6 -- 36.3\\
			~ & $\theta^{}_{13}$ $[{}^{\circ}_{}]$ & 8.60 & 8.48 -- 8.74 & 8.22 -- 8.99 \\
			~ & $\theta^{}_{23}$ $[{}^{\circ}_{}]$& 48.8 & 46.4 -- 49.4 & 41.7 -- 51.3\\ 
			~ & $\delta$ $[{}^{\circ}_{}]$& 285 & 280 -- 289 & 90 -- 298\\
			~ & $\rho$ $[{}^{\circ}_{}]$& 89.5 & 89.3 -- 89.8 & 0 -- 180\\
			~ & $\sigma$ $[{}^{\circ}_{}]$& 90.6 & 90.2 -- 90.7 & 0 -- 180\\
			~ & $m^{}_{\beta}$ $[{\rm eV}]$& 0.1070 & 0.09602 -- 0.1834 & 0.04867 -- 0.6592\\
			~ & $m^{}_{\beta\beta}$ $[{\rm eV}]$ & 0.1066 & 0.09514 -- 0.1847 & 0.04526 -- 0.6509\\
			\hline
			\hline
		\end{tabular}
		\label{table:value}
	\end{center}
\end{table}

As a summary of this section, we list the best-fit values, together with the $1\sigma$ and $3\sigma$ ranges of all the free model parameters and observables in our model in Table~\ref{table:value}.

\section{Summary} \label{sec:summary}
The modular symmetry is a very attractive and interesting way to understand lepton flavor mixing. In this paper, we consider the application of the modular $A^{}_4$ symmetry to the supersymmetric minimal type-(I+II) seesaw model, where only one right-handed neutrino and two Higgs triplets are introduced beyond the particle content of the SM. We successfully construct a model to account for lepton mass spectra and the flavor mixing, which is consistent with current neutrino oscillation data in the NO case.

In order to keep our model simple and economical enough, we assign the right-handed neutrino, two Higgs doublets and two Higgs triplets to be the trivial $A^{}_{4}$ singlet {\bf 1}, and implement the minimal set for the weights of modular forms $(k^{}_{e}, k^{}_{\mu}, k^{}_{\tau}, k^{}_{\Delta}, k^{}_{\rm D}, k^{}_{\rm R}) = (2,2,2,2,4,6)$ under the premise that the sum of weights in each superpotential should be vanishing. We construct the mass matrices in both the charged-lepton and neutrino sectors under such a setup of weights. After performing the numerical analysis, we find out that our model is consistent with the latest global-fit results of neutrino oscillation data at the $1\sigma$ level only in the NO case and the individual contributions to the neutrino masses from the right-handed neutrino and the Higgs triplet are comparable. The allowed parameter space of the model parameters, namely, the modulus parameter $\tau = {\rm Re}\,\tau + {\rm i}\,{\rm Im}\,\tau$, three real parameters $v^{}_{\rm d} \alpha^{}_3 /\sqrt{2}$, $\alpha^{}_1/\alpha^{}_3$ and $\alpha^{}_2/\alpha^{}_3$ in the charged-lepton sector, together with the coupling coefficient $\widetilde{g} = g e^{{\rm i}\phi^{}_g}$ and the overall scale $v^2_{\rm u} g^2_1/(2\Lambda)$ in the neutrino sector has been obtained. Moreover, we also give the constrained regions of three light neutrino masses $\{m^{}_1,m^{}_2,m^{}_3\}$, three neutrino mixing angles $\{\theta^{}_{12}, \theta^{}_{13}, \theta^{}_{23}\}$ and three CP-violating phases $\{\delta, \rho, \sigma\}$, as well as the predictions for the effective neutrino masses $m^{}_{\beta}$ in beta decays and $m^{}_{\beta\beta}$ in neutrinoless double-beta decays.

An interesting feature of our model is that the octant of $\theta^{}_{23}$ strongly depends on which hierarchy of $\alpha^{}_1$, $\alpha^{}_2$ and $\alpha^{}_3$ is taken into consideration. To be specific, the hierarchy $\alpha^{}_1 \ll \alpha^{}_3 \ll \alpha^{}_2$ corresponds to the first octant of $\theta^{}_{23}$ while the hierarchy $\alpha^{}_1 \ll \alpha^{}_2 \ll \alpha^{}_3$ corresponds to the second octant of $\theta^{}_{23}$. Furthermore, when the value of ${\rm Im}\,\tau$ is sufficiently large, the neutrino mass matrix $M^{}_{\nu}$ will possess an approximate $\mu-\tau$ reflection symmetry, indicating $\theta^{}_{23} \approx 45^\circ_{}$ and $\delta \approx  90^{\circ}_{}$ or $270^{\circ}_{}$. While the small parameter $t \equiv -6q^{1/3}_{}= -6 e^{2\pi{\rm i}\tau/3}$ slightly breaks down this symmetry and generate realistic values for the mass-squared differences and other mixing angles. Besides, since our model predicts relatively large values of $\sum m^{}_{\nu} =m^{}_1+m^{}_2+m^{}_3$, $m^{}_{\beta}$ and $m^{}_{\beta\beta}$, it is very likely to be tested in the further neutrino experiments.

We stress that the hybrid seesaw models where not only one kind of seesaw mechanism is involved may lead to some new scenarios about flavor mixing and CP violation, and it deserves more attention to discuss the applications of the modular symmetry in such kinds of models. It is also interesting to study the type-(I+II) seesaw model with other kinds of finite modular symmetries. We hope to come back to these issues in the future works.

\section*{Acknowledgements}

I am greatly indebted to Prof.~Shun Zhou for suggesting this work and carefully reading this manuscript. I would also like to thank Guo-yuan Huang, Dr.~Biswajit Karmakar and Di Zhang for helpful discussions. This work was supported in part by the National Natural Science Foundation of China under grant No.~11775232 and No.~11835013.

\appendix

\section{The $\Gamma^{}_{3} \simeq A^{}_{4}$ Symmetry Group}\label{sec:appA}
The permutation symmetry group $A^{}_{4}$ has twelve elements and four irreducible representations, which are denoted as ${\bf 1}$, ${\bf 1^{\prime}_{}}$, ${\bf 1^{\prime\prime}_{}}$ and ${\bf 3}$. In the present work, we choose the complex basis which is used in Ref.~\cite{Feruglio:2017spp} for the representation matrices of two generators $S$ and $T$, namely,
\begin{eqnarray}
\begin{array}{cclcl}
{\bf 1} &: ~\quad & \rho(S) =  1 \; , & ~\quad & \rho(T) = 1 \; , \\
{\bf 1^{\prime}_{}} &: ~\quad & \rho(S)=1 \; , & ~\quad & \rho(T)=\omega \; , \\
{\bf 1^{\prime}_{}} &: ~\quad & \rho(S)=1 \; , & ~\quad & \rho(T)=\omega^2_{} \; , \\
{\bf 3} &: ~\quad & \displaystyle \rho(S)=\frac{1}{3}\left(\begin{matrix}
-1 && 2 && 2 \\
2 && -1 && 2 \\
2 && 2 && -1
\end{matrix}\right) \; , &~\quad & \rho(T)=\left(\begin{matrix}
1 && 0 && 0 \\
0 && \omega && 0 \\
0 && 0 && \omega^2_{}
\end{matrix}\right) \; .
\end{array}
\label{eq:A4rep}
\end{eqnarray}
In this basis, we can explicitly write down the decomposition rules of the Kronecker products of any two $A^{}_{4}$ multiplets.
\begin{itemize}
	\item For the Kronecker products of two $A^{}_4$ singlets:
	\begin{align}
	{\bf 1}\otimes {\bf 1} = {\bf 1} \; ,  \quad 
	{\bf 1^{\prime}_{}} \otimes {\bf 1^{\prime}_{}} = {\bf 1^{\prime\prime}_{}} \; ,  \quad
	{\bf 1^{\prime\prime}_{}} \otimes {\bf 1^{\prime\prime}_{}} = {\bf 1^{\prime}_{}} \; ,  \quad
	{\bf 1^{\prime}_{}} \otimes {\bf 1^{\prime\prime}_{}} = {\bf 1} \; ; 
	\label{eq:1p3p}
	\end{align}
	\item For the Kronecker products of two $A^{}_4$ triplets:
	\begin{eqnarray}
	\left(\begin{matrix}
	\zeta^{}_{1} \\ \zeta^{}_{2} \\ \zeta^{}_{3}
	\end{matrix}\right)^{}_{\bf 3} \otimes
	\left(\begin{matrix}
	\xi^{}_{1} \\ \xi^{}_{2} \\ \xi^{}_{3}
	\end{matrix}\right)^{}_{\bf 3}  
	= && (\zeta^{}_{1}\xi^{}_{1}+\zeta^{}_{2}\xi^{}_{3}+\zeta^{}_{3}\xi^{}_{2})^{}_{\bf 1} \oplus (\zeta^{}_{3}\xi^{}_{3}+\zeta^{}_{1}\xi^{}_{2}+\zeta^{}_{2}\xi^{}_{1})^{}_{\bf 1^{\prime}_{}} \nonumber \\ 
	&& \oplus  (\zeta^{}_{2}\xi^{}_{2}+\zeta^{}_{1}\xi^{}_{3}+\zeta^{}_{3}\xi^{}_{1})^{}_{\bf 1^{\prime\prime}_{}} \nonumber  \\
	&& \oplus \dfrac{1}{3}\left(\begin{matrix}
	2\zeta^{}_{1}\xi^{}_{1}-\zeta^{}_{2}\xi^{}_{3}-\zeta^{}_{3}\xi^{}_{2} \\ 2\zeta^{}_{3}\xi^{}_{3}-\zeta^{}_{1}\xi^{}_{2}-\zeta^{}_{2}\xi^{}_{1} \\
	2\zeta^{}_{2}\xi^{}_{2}-\zeta^{}_{1}\xi^{}_{3}-\zeta^{}_{3}\xi^{}_{1}
	\end{matrix}\right)^{}_{\bf 3}
	\oplus \frac{1}{2}\left(\begin{matrix}
	\zeta^{}_{2}\xi^{}_{3}-\zeta^{}_{3}\xi^{}_{2} \\ \zeta^{}_{1}\xi^{}_{2}-\zeta^{}_{2}\xi^{}_{1} \\
	\zeta^{}_{3}\xi^{}_{1}-\zeta^{}_{1}\xi^{}_{3}
	\end{matrix}\right)^{}_{\bf 3} \; .  \label{eq:33} 
	\end{eqnarray}
\end{itemize}

With the above decomposition rules and the assignments of relevant fields and modular forms, one can immediately find out the Lagrangian invariant  under the modular $A^{}_4$ symmetry group.

As has been mentioned in Sec.~\ref{sec:modular}, there exist three linearly independent modular forms of the lowest non-trivial weight $k^{}_{Y}=2$, denoted as $Y^{}_i(\tau)$ for $i = 1, 2, 3$. They transform as a triplet ${\bf 3}$ under the $A^{}_4$ symmetry, namely,
\begin{eqnarray}
Y^{}_{\bf 3} (\tau) \equiv  \left(\begin{matrix} Y^{}_{1}(\tau) \\ Y^{}_2 (\tau) \\ Y^{}_{3} (\tau) \end{matrix}\right) \; .
\label{eq:S4Y1}
\end{eqnarray}
In fact, the exact expressions of the modular forms can be derived with the help of the Dedekind $\eta$ function
\begin{eqnarray}
\eta(\tau) \equiv q^{1/24}_{} \prod_{n=1}^{\infty}(1-q^{n}_{}) \; ,
\label{eq:eta}
\end{eqnarray}
with $q \equiv e^{2\pi {\rm i} \tau}_{}$, and its derivative
\begin{equation}
\begin{split}
Y(a^{}_{1}, \dots ,a^{}_{4}|\tau) \equiv
\frac{\rm d}{\rm{d}\tau}\bigg[ a^{}_{1} \log \eta \left(\frac{\tau}{3} \right) + a^{}_{2} \log \eta \left(\frac{\tau+1}{3}\right) + a^{}_{3} \log \eta \left(\frac{\tau+2}{3} \right) + a^{}_{4} \log \eta \left(3\tau \right) \bigg] \; ,
\end{split}
\label{eq:geneY}
\end{equation}
with the coefficients $a^{}_i$ (for $i = 1, 2, \cdots, 4$) fulfilling $a^{}_{1}+ \cdots +a^{}_{4} = 0$. More explicitly, we have
\begin{eqnarray}
Y^{}_{1}(\tau)  & \equiv & \frac{\rm i}{2\pi} \left[\frac{\eta^{\prime}_{}\left(\frac{\tau}{3}\right)}{\eta 
\left(\frac{\tau}{3}\right)}+\frac{\eta^{\prime}_{}\left(\frac{\tau+1}{3}\right)}{\eta 
\left(\frac{\tau+1}{3}\right)}+\frac{\eta^{\prime}_{}\left(\frac{\tau+2}{3}\right)}{\eta 
\left(\frac{\tau+2}{3}\right)}-\frac{27\eta^{\prime}_{}\left(3\tau\right)}{\eta 
\left(3\tau\right)}\right] \; , \nonumber \\
Y^{}_{2}(\tau)  & \equiv & \frac{-\rm i}{\pi} \left[\frac{\eta^{\prime}_{}\left(\frac{\tau}{3}\right)}{\eta 
	\left(\frac{\tau}{3}\right)}+\omega^2_{}\frac{\eta^{\prime}_{}\left(\frac{\tau+1}{3}\right)}{\eta 
	\left(\frac{\tau+1}{3}\right)}+\omega\frac{\eta^{\prime}_{}\left(\frac{\tau+2}{3}\right)}{\eta 
	\left(\frac{\tau+2}{3}\right)}\right] \; , \nonumber \\
Y^{}_{3}(\tau)  & \equiv & \frac{-\rm i}{\pi} \left[\frac{\eta^{\prime}_{}\left(\frac{\tau}{3}\right)}{\eta 
	\left(\frac{\tau}{3}\right)}+\omega\frac{\eta^{\prime}_{}\left(\frac{\tau+1}{3}\right)}{\eta 
	\left(\frac{\tau+1}{3}\right)}+\omega^2_{}\frac{\eta^{\prime}_{}\left(\frac{\tau+2}{3}\right)}{\eta 
	\left(\frac{\tau+2}{3}\right)}\right] \; ,
\label{eq:Yexp}
\end{eqnarray}
which can be expanded as the Fourier series, i.e.,
\begin{eqnarray}
Y^{}_{1}(\tau) &=& 1+12q+36q^2_{}+12q^3_{}+\cdots \; , \nonumber  \\
Y^{}_{2}(\tau) &=& -6q^{1/3}_{}(1+7q+8q^2_{}+\cdots) \; , \nonumber  \\
Y^{}_{3}(\tau) &=& -18q^{2/3}_{}(1+2q+5q^2_{}+\cdots) \; . \label{eq:Y3q} 
\end{eqnarray}

\end{document}